\documentclass[preprint2]{proto}
\usepackage{times}
\newcommand{\refs}{\par\noindent\hangindent=1pc\hangafter=1}
\begin{document}

\title{\textbf{\LARGE An Observational Perspective of Low-Mass Dense Cores I: \\
Internal Physical and Chemical Properties}}

\author {\textbf{\large James Di Francesco}}
\affil{\small\em National Research Council of Canada}
\author {\textbf{\large Neal J. Evans II}}
\affil{\small\em The University of Texas at Austin}
\author {\textbf{\large Paola Caselli}}
\affil{\small\em Arcetri Observatory}
\author {\textbf{\large Philip C. Myers}}
\affil{\small\em Harvard-Smithsonian Center for Astrophysics}
\author {\textbf{\large Yancy Shirley}}
\affil{\small\em University of Arizona}
\author {\textbf{\large Yuri Aikawa}}
\affil{\small\em Kobe University}
\author {\textbf{\large Mario Tafalla}}
\affil{\small\em National Astronomical Observatory of Spain}

\begin{abstract}
\baselineskip = 11pt
\leftskip = 0.65in 
\rightskip = 0.65in
\parindent=1pc
{\small 
Low-mass dense cores represent the state of molecular gas associated with
the earliest phases of low-mass star formation.  Such cores are called
``protostellar" or ``starless," depending on whether they do or do not
contain compact sources of luminosity.  In this chapter, the first half 
of the review of low-mass dense cores, we describe the numerous inferences 
made about the nature of starless cores as a result of recent observations, 
since these reveal the initial conditions of star formation.  We focus on 
the identification of isolated starless cores and their internal physical 
and chemical properties, including morphologies, densities, temperatures, 
kinematics, and molecular abundances.  These objects display a wide range 
of properties since they are each at different points on evolutionary paths 
from ambient molecular cloud material to cold, contracting, and centrally 
concentrated configurations with significant molecular depletions and, in 
rare cases, enhancements.
 \\~\\~\\~}

\end{abstract}  

\bigskip
\section{\textbf{INTRODUCTION AND SCOPE OF REVIEW}}
\medskip

Over the last decade, dedicated observations have revealed much about the 
earliest phases of low-mass star formation, primarily through studies of 
low-mass dense cores with and without internal protostellar sources, e.g.,
protostellar and starless cores.  Such cores are dense zones of molecular 
gas of relatively high density, and represent the physical and chemical
conditions of interstellar gas just after or prior to localized gravitational 
collapse.  These objects are essentially the seeds from which young stars 
may spring, and define the fundamental starting point of stellar evolution.  
Since PPIV (see {\it Langer et al.}\ 2000; {\it Andr\'e et al.}\ 2000), 
enormous strides have been made in the observational characterization of 
such cores, thanks to the increased interest stemming from the promising 
early results reported at PPIV and to new instrumentation.

This review summarizes the advances made since PPIV, and, given the large 
number, it has been divided into two parts.  In Part I (i.e., this chapter), 
we summarize major recent observational studies of the initial conditions 
of star formation, i.e., the individual physical and chemical properties of 
starless cores, with emphasis on cores not found in crowded regions.  This 
subject includes their identification, morphologies, densities, temperatures, 
molecular abundances, and kinematics.  In recent years, the internal density 
and temperature structures of starless cores have been resolved, regions of 
chemical depletion (or enhancement) have been studied, and inward motions have 
been measured.  Part II of this review (see next chapter by {\it Ward-Thompson 
et al.}) summarizes what observed individual and group characteristics, 
including magnetic field properties, mass distributions and apparent lifetimes, 
have taught us about the evolution of low-mass dense cores through protostellar 
formation.  Despite significant progress, the observational picture of the 
earliest stages of star formation remains incomplete, and new observational
and experimental data and theoretical work will be necessary to make further 
advances.

\medskip
\section{\textbf{STARLESS CORES}}
\medskip

\noindent
\textbf{2.1 Definition}
\medskip

Stars form within molecular gas behind large amounts of extinction from dust.  
This extinction has been used to locate molecular clouds ({\it Lynds}, 1962), 
isolated smaller clouds ({\it Clemens and Barvainis}, 1988), and particularly 
opaque regions within larger clouds ({\it Myers et al.}, 1983; {\it Lee and 
Myers}\ 1999).  Locations within these clouds of relatively high density or 
column density, though identifiable from optical or infrared absorption (e.g., 
see chapter by {\it Alves et al.}), can also be detected using 
submillimeter, millimeter, or radio emission.  For instance, 
spectroscopic studies of clouds found via extinction studies further selected 
those with evidence for dense gas (e.g., see {\it Myers et al.}, 1983; {\it 
Myers and Benson}\ 1983).  These became known collectively as ``dense cores," 
based primarily on whether or not they showed emission 
from NH$_{3}$, indicative of densities above about $10^{4}$ cm$^{-3}$ ({\it
Benson and Myers}, 1989; see also {\it Jijina et al.}, 1999).  Using 
IRAS data, {\it Beichman et al.}\ (1986) found that roughly half the known 
dense cores in clouds contained IRAS sources, while {\it Yun and Clemens}\ 
(1990) found that about 23\% of the isolated smaller clouds contained 
infrared sources within; the remainder became known as ``starless cores."  
Follow-up studies of clouds with sensitive submillimeter continuum detection 
systems allowed further discrimination within the class of starless cores.  
Those with detected emission, indicating relatively high densities of 
$10^{5-6}$ cm$^{-3}$ ({\it Ward-Thompson et al.}, 1994), were called 
``pre-protostellar" cores or later ``prestellar cores."  Molecular line 
studies showed that prestellar cores were indeed more likely to show evidence 
for inflowing material than were the merely starless cores (e.g., see {\it
Gregersen and Evans}, 2000).
 
In this review, we describe the recent observational characterization of
``starless cores," low-mass dense cores without compact luminous sources of 
any mass.  (We call such sources ``young stellar objects," regardless of 
whether or not they are stellar or substellar in mass.)  By ``low-mass," we 
mean cores with masses $M< 10$ M$_{\odot}$, such as those found in the nearby 
Taurus or Ophiuchus clouds.  Starless cores are important because they 
represent best the physical conditions of dense gas prior to star formation.  
Conceptually, we distinguish starless cores that are gravitationally bound as 
``prestellar cores" though it is very difficult to determine observationally 
at present whether a given core is bound or not (see below).  In addition, we 
consider such cores as either ``isolated" or ``clustered" depending on the 
local surface density of other nearby cores and young stellar objects.  
Observations of isolated cores (e.g., {\it Clemens and Barvainis}, 1988) 
are easier to interpret since they generally inhabit simpler, less confused 
environments, i.e., without nearby cores or young stellar objects.  
Observations of clustered cores (e.g., {\it Motte et al.}, 1998) 
are important also because they are more representative of star formation 
within embedded clusters, i.e., where most stars form within the Galaxy ({\it
Lada and Lada}, 2003).  Finally, we describe cores as either ``shielded" from
or ``exposed" to the interstellar radiation field (ISRF), depending on whether 
they do [e.g., many of those described by {\it Myers et al.}, 1983] or do not 
[e.g., those cataloged by {\it Clemens and Barvainis}, 1988] reside within a 
larger molecular cloud.

Observational limitations strongly influence how dense cores are defined.  
For example, low resolution data may reveal cores over a wide range of mass, 
but higher resolution data may reveal substructure within each, with their own 
range of (lower) mass, i.e., cores within cores.  In addition, different 
tracers may probe different kinds of structure within clouds, due to varying 
density, temperature, or chemistry.  Also, whether or not a given dense 
core contains a compact infrared-luminous source is ultimately a matter of 
observational sensitivity (e.g., L1014; see below).  Finally, learning
whether or not a given core is prestellar (i.e., gravitationally bound) 
requires accurate determination of its physical state (mass, temperature, 
and internal magnetic field).  Masses alone can be uncertain by factors 
of $\sim$3 due to uncertainties in distances, dust opacities, molecular 
abundances, or calibration.  Various new instrumentation of higher resolution 
and sensitivity will soon be available, making such observational 
characterization more tenable.

\begin{figure*}
\epsscale{1.0}
\plotone{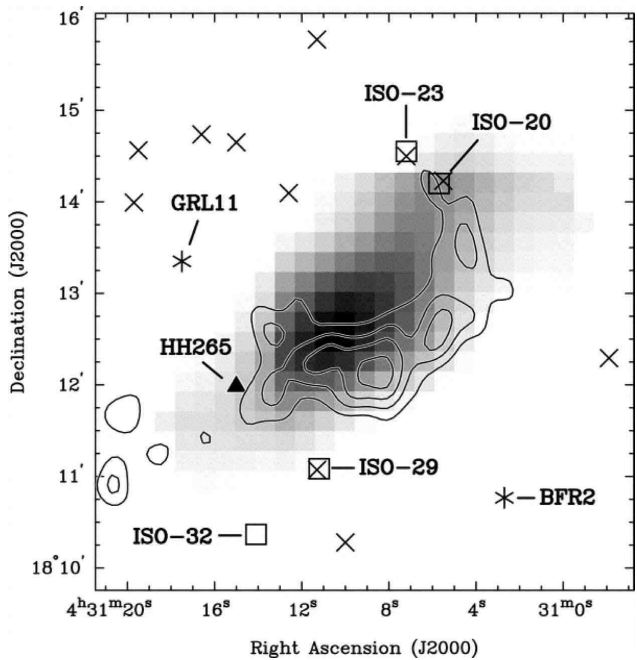}
\caption{\small The prestellar core L1551-MC shown in NH$_{3}$ (1,1) 
integrated intensity (greyscale) and CCS(3$_{2}$--2$_{1}$) integrated 
intensity (contours).  Positions of nearby objects are shown as symbols 
(see {\it Swift et al.}, 2005).}
\end{figure*}

\medskip
\noindent
\textbf{2.2 Identification}
\medskip

Observations of starless cores have focused mostly on examples within 1 kpc, 
where the highest mass sensitivities and linear resolutions are possible.  
Many starless cores have been detected serendipitously, adjacent to 
previously-known protostellar cores or young stellar objects (e.g., see {\it 
Wilson et al.}, 1999; {\it Williams and Myers}, 1999; {\it Swift et al.},
2005).  Only recently, more systematic mapping has begun to identify larger 
numbers of starless cores in less biased surveys.

Recent large line maps across significant portions of nearby molecular clouds 
using transitions that trace relatively high densities have been useful in 
identifying dense cores.  For example, 179 cores in clouds within 200 pc were 
identified from substructure detected in C$^{18}$O (1--0) maps of L1333 in 
Cassiopeia ({\it Obayashi et al.}, 1998), Taurus ({\it Onishi et al.}, 1998), 
the Southern Coalsack region ({\it Kato et al.}, 1999), the Pipe Nebula region 
({\it Onishi et al.}, 1999), Chamaeleon ({\it Mizuno et al.}, 1999), Lupus 
({\it Hara et al.}, 1999), Corona Australis ({\it Yonekura et al.}, 1999), 
and Ophiuchus ({\it Tachihara et al.}, 2000).  {\it Tachihara et al.}\ (2002) 
identified 76\% of these cores as ``starless" (i.e., without associated IRAS 
sources; see below) with masses of 1-100 M$_{\odot}$ (on average, 15 
M$_{\odot}$).  Using a still higher density molecular tracer, H$^{13}$CO$^{+}$ 
(1--0), {\it Onishi et al.}\ (2002) identified 55 
``condensations" in Taurus, 80\% of which were similarly deemed ``starless" 
with estimated masses of 0.4-20 M$_{\odot}$.   Using line maps to identify 
cores can be problematic, however, because of line optical depths and chemical 
evolution within the cloud.  For example, lines like C$^{18}$O (1--0) can be 
too opaque to sample most of the material in dense cores.  Furthermore, this 
line and H$^{13}$CO$^{+}$ (1--0) can be affected by the significant depletion 
of the source molecules, again making it difficult to sample their column 
densities.  In contrast, N--bearing molecules, in particular NH$_3$ and 
N$_2$H$^+$, appear to survive at higher densities than C-bearing species, 
and in fact they have been used to probe core nuclei (see {\it Caselli et 
al.}, 2002c; also Section 4).
 
Submillimeter or millimeter continuum emission maps of molecular clouds can 
trace the high column densities associated with starless cores without the 
potentially confusing effects of line opacity or chemical evolution.  These 
studies trace the dust component, which emits radiation at long wavelengths 
even at the very low dust temperatures ($T_D$) in dense cores ($\sim$10 K), 
though the emission is sensitive to $T_D$.  The optical properties of grains 
are also likely to evolve at relatively high densities due to ice mantle 
growth and coagulation, but these effects are much less severe than those 
for gas phase tracers (see Section 4).  Recent single-dish telescope studies 
include the inner regions of Ophiuchus ({\it Motte et al.}, 1998; {\it 
Johnstone et al.}, 2000; {\it Johnstone et al.}, 2004; {\it Stanke et al.}, 
2006), northern Orion A ({\it Chini et al.}, 1997; {\it Lis et al.}, 1998; 
{\it Johnstone and Bally}, 1999), NGC 2068/2071 in Orion B ({\it Mitchell 
et al.}, 2001; {\it Johnstone et al.}, 2001; {\it Motte et al.}, 2001), NGC 
1333 in Perseus ({\it Sandell and Knee}, 2001), and R Corona Australis ({\it 
Chini et al.}, 2003; {\it Nutter et al.}, 2005).  For example, {\it Motte et
al.}\ (1998) found 58 ``starless clumps" in their map of Ophiuchus with masses 
of 0.05-3 M$_{\odot}$.  Even larger regions (e.g., 8--10 deg$^2$) have now 
been mapped ({\it Enoch et al.}, 2006; {\it Young et al.}, 2006; {\it Hatchell 
et al.}, 2005; {\it Kirk et al.}, 2006), providing expansive (though shallow) 
coverage of the Perseus and Ophiuchus clouds.  Collectively these studies 
have revealed hundreds of compact continuum objects.  In those cases where 
comparisons with near- to far-infrared data were made, more than $\sim$60\% 
were deemed ``starless."  The fraction of these that are prestellar is not 
known, however, and determining this (through appropriate molecular tracers; 
see Section 4) is a key goal of future research. 

In both line and continuum maps, the methods by which objects are identified
have varied widely.  Some studies have identified objects by eye, while others 
have used more objective automated structure identification algorithms (e.g.,
{\it Clumpfind} by {\it Williams et al.}, 1994), multiscale wavelet analyses, 
and even hybrid techniques.  In addition, objects identified in continuum 
studies are significantly more compact in scale than those found in the line 
maps described above.  First, angular resolutions of the data have
substantially differed, with line and continuum maps typically having $>$ 
1\arcmin\ and $<$ 1\arcmin\ resolutions respectively.  Second, the spatial 
sampling of the data also have substantially differed, though single-dish 
telescopes were used for both.  Line data are obtained through frequency or 
position switching to line-free locations, allowing large-scale information 
to be retained, but continuum data are obtained through chopping (or sky 
background removal for instruments that do not chop), filtering out large-scale 
information (e.g., over 2--10\arcmin\ depending on the instrument.)  Indeed, 
many continuum objects can be found embedded within objects identified via 
line surveys.  Continuum cores have been identified, however, using relatively 
low-resolution data, e.g., from IRAS ({\it Jessop and Ward-Thompson}, 2000) or 
ISO ({\it T\'oth et al.}, 2004).
  
Identification of objects found through line or continuum observations as 
starless cores is ultimately dependent upon the availability and sensitivity 
of ancillary data at the time of analysis.  For example, Figure 1 shows the 
starless core L1551-MC in Taurus recently discovered by {\it Swift et al.}\
(2005) in NH$_{3}$ and CCS emission, with the positions of nearby, 
non-coincident objects listed in the SIMBAD database (as of late 2004).  
Far-infrared data from IRAS were used traditionally to find evidence 
for embedded protostars associated with detected cores, but more recently 
wide-field near-infrared data, including those from 2MASS, have been also 
used.  Such wide-field data can be insufficiently sensitive to detect the 
youngest, most-embedded protostars, however, given high extinctions (e.g., 
$A_{V}$ $>$ 30).  Near-infrared observations targeting specific cores can 
attain much higher sensitivities but these have not been done extensively. 
Some targeted studies, e.g., {\it Allen et al.}\ (2002) or {\it Murphy and 
Myers}\ (2003), however, did not detect low luminosity protostars in various 
cores despite relatively high sensitivity.  The extraordinarily high 
sensitivity of the Spitzer Space Telescope in the mid-infrared has revealed 
about 12 low luminosity protostars within cores previously identified as 
starless, e.g., in L1014 by {\it Young et al.} (2004a) and in L1148 by {\it 
Kauffmann et al.} (2005; (see also {\it Bourke et al.}, in preparation; {\it 
Huard et al.}, in preparation).  These are candidates to join the class 
of Very Low Luminosity Objects (VeLLOs), sources with L$_{int}$ $<$ 0.1 
L$_{\odot}$ that are located within dense cores.  (Note that L$_{int}$ is 
the luminosity of the object, in excess of that supplied by the ISRF.)  
Based on this definition, the previously identified, extremely young, 
low-luminosity Class 0 protostar IRAM 04191 ({\it Andr\'e et al.}, 1999) 
is also a VeLLO ({\it Dunham et al.}, in preparation).
 
In addition to near- to far-infrared imaging, deeply embedded protostars may 
be identified using other signposts of star formation not affected by high 
extinction, including compact molecular outflows, ``hot core" molecular line 
emission, masers, or compact thermal radio emission.  For example, {\it Yun 
et al.}\ (1996) and {\it Harvey et al.}\ (2002) used the latter method to 
limit luminosities of protostellar sources within several starless cores 
using limits on radio emission expected from shock-ionized gas, i.e., from 
outflows.  

\medskip
\section{\textbf{DENSITY AND TEMPERATURE STRUCTURES}}
\medskip

\noindent
\textbf{3.1 Probing Physical Structure}
\medskip

In this section, we describe the morphologies and internal configurations 
of density and temperature of starless cores as derived in recent studies.  
Much progress in understanding dense core structure has come from analyses 
of far-infrared to millimeter continuum emission observations, typically of 
relatively isolated and exposed examples.  Continuum data have been used 
preferentially over molecular line data because interpretation of the latter 
can be complicated by large internal molecular abundance variations (see 
Section 4).  Near- or mid-infrared absorption of background emission by cores 
can be also a powerful probe of the density structure since extinction is 
independent of temperature and can trace those locations where dust column 
densities are too low to be easily detected in emission at present (e.g., 
see {\it Bacmann et al.}, 2000).  Results about core structure derived 
from ``extinction mapping" have agreed generally with those from emission 
studies, with some exceptions (see below and the chapter by {\it Alves et 
al.})  

The observed emission from cores depends on the entangled effects of
temperature, density, properties of the tracer being used, and the 
observational technique. For molecular lines, the abundance and excitation
of the tracer varies within the core. For continuum emission from dust,
the interpretation depends on dust properties and the characteristics of
the ISRF. Two general approaches are taken: in the first, simplifying 
assumptions are made and the observations are inverted to derive core 
properties, usually averages over the line of sight and the beam; in 
the second, specific physical models are assumed and a self-consistent 
calculation of predicted observations is compared to the actual 
observations, allowing the best model to be chosen.  Both approaches 
have their strengths and weaknesses.

Self-consistent models must compute the temperature distribution for 
a given density distribution, dust properties, and radiation field.
For starless cores, only the ISRF heats the dust.  In their models, 
{\it Evans et al.}\ (2001) included the ``Black-Draine" (BD) ISRF, one based 
on the COBE results of {\it Black}\ (1994), which includes the cosmic microwave 
background (CMB) blackbody and the ISRF in the ultraviolet of {\it Draine}\ 
(1978; cf. {\it van Dishoeck}\ 1988), rather than the older, pre-COBE model 
of {\it Mathis et al.}\ (1983; MMP).  The BD and MMP models can 
differ significantly, e.g., up to a factor of 13 between 5 $\mu$m and 400 
$\mu$m.  For dust opacities, those currently in use have been theoretically
derived from optical constants of grain and ice materials and include features
such as mid-infrared vibrational bands (e.g., $9.7$ $\mu$m Si-O stretch) and
tend toward a power-law decrease of opacity at submillimeter wavelengths
($\kappa_{\nu} \propto \lambda^{-\beta}$).  Figure 2a shows five dust opacity
models: OH5 and OH8 are models of grains that have coagulated for $10^5$ yr
and acquired varied depths of ice mantles ({\it Ossenkopf and Henning}, 1994); 
``Draine-Lee" are the ``standard'' interstellar medium dust opacities derived 
from silicate and graphite grains by {\it Draine and Lee}\ (1984); MMP 
opacities are derived from an empirical fit to observed dust properties in 
star-forming regions; and Pollack opacities are derived from an alternative 
grain composition based on silicate grains (olivine and orthopyroxene), iron 
compounds (troilite and metallic iron), and various organic C compounds ({\it
Pollack et al.}, 1994).  The mass opacity ($\kappa_{\nu}$) at long wavelengths 
can vary by one order of magnitude between the opacity models.

\begin{figure*}
\epsscale{2.0}
\plotone{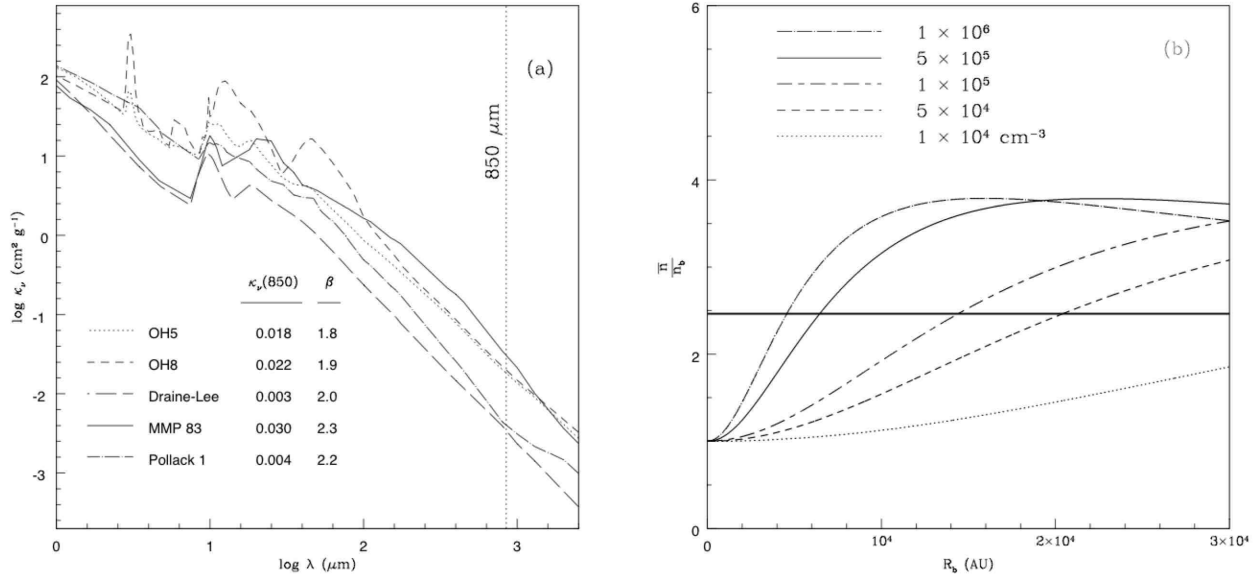}
\caption{\small
a) Opacity vs. wavelength from selected dust models (see Section 3.3; {\it 
Shirley et al.}, 2005).  Values ($\kappa_{\nu}$) and spectral indices ($\beta$) 
of opacities at 850 $\mu$m are tabulated.  Units of $\kappa_{\nu}$ are cm$^{2}$ 
g$^{-1}$ and the $\beta$ at 850 $\mu$m were obtained from linear regressions 
of respective model values between 350 $\mu$m and 1.3 mm.  b) The ratio of 
average density ($\bar n$) to density at the BES boundary radius ($n_b$) vs. 
BES boundary radius ($R_b$) for families of BES of given $n_c$ ({\it Shirley 
and Jorgensen}, in preparation).  The solid horizontal line corresponds to the 
stability criterion of {\it Lombardi and Bertin}\ (2001).} 
\end{figure*}

These ingredients then provide inputs to radiative transport codes (e.g., 
{\it Egan et al.}, 1988; {\it Ivezi\'{c} et al.,}, 1999), which attempt 
to fit simultaneously the observed intensity profiles, $I_{\nu}(r)$ (most 
sensitive to variation in $n(r)$), and spectral energy distribution (SED, 
most sensitive to the total mass and the assumed dust opacity).  To test 
various ISRF and dust opacity models for the L1498 starless core, {\it 
Shirley et al.}\ (2005) varied the BD ISRF by a scaling 
factor, $s_{isrf}$, and tested eight dust opacity models, including those 
shown in Figure 2a.  In their models, the resultant SED was more sensitive 
to the ISRF than the resultant normalized continuum intensity profiles, and 
they found best fits with $s_{isrf}$ = 0.5--1.0.  Furthermore, only the OH5 
and OH8 opacity models were consistent with the observed SED.

Chemical differentiation observed in starless cores (see Section 4) should 
lead to variations in ice compositions, and hence dust opacities.  Indeed, 
{\it Kramer et al.}\ (2003) found in IC 5146 an increase in dust emissivity by 
a factor of 4 between 20 K and 12 K.  To date, variations of dust opacity 
with radius have not been self-consistently included in radiative transfer 
models.  Moreover, most radiative transfer models are 1D and are constrained 
by 1D intensity profiles, although  starless cores can be quite ellipsoidal 
in 3D shape (see \S3.2).  If magnetic fields are dynamically important, they 
require core geometries that are inherently multi-dimensional since the 
magnetic field only provides support perpendicular to the direction of the 
field lines (e.g., {\it Li and Shu}, 1996).  Promising work on modeling 
cores with 3D radiative transfer codes has begun (e.g., {\it Doty et al.}, 
2005a, 2005b; {\it Steinacker et al.}, 2005).  A reanalysis of L1544 core 
data by {\it Doty et al.}\ (2005a) using a 3D code yielded results that were 
overall similar to those found from 1D models but with reduced size and mass.  

\medskip
\noindent
\textbf{3.2  Morphology and Density Structure}
\medskip

Starless cores vary widely in morphology or shape, ranging between filamentary 
to compact and round.  Cores are seen as 2D projections against the sky, making 
it difficult to constrain their 3D shape.  Furthermore, core morphologies can 
depend on the wavelength and the angular resolution of observation (e.g., see 
{\it Stamatellos et al.}, 2004).  Observed morphologies also depend upon the 
intensity level chosen as the boundary between cores and their surroundings, 
and this can be affected strongly by observational sensitivity.  In general, 
cores that are round (in projection) have been typically interpreted as being
spherical in shape, while cores that are elongated (in projection) have been 
typically interpreted as being either oblate or prolate spheroids.  Statistical
analyses of core map aspect ratios indicate that most isolated cores are 
prolate rather than oblate ({\it Myers et al.}, 1991; {\it Ryden}, 1996), and 
many cores in Taurus share the projected alignment of the filamentary clouds 
in which they reside ({\it Hartmann}, 2002).  {\it Jones et al.}\ (2001), 
using a sample of cores identified from the extensive NH$_{3}$ catalog of 
{\it Jijina et al.}\ (1999), posited that elongated cores actually have 
intrinsically triaxial shapes that are neither purely oblate nor prolate.  
{\it Goodwin et al.}\ (2002), using the same catalog, found starless cores 
have more extreme axial ratios than protostellar cores.

Column densities within starless cores can be traced using the optically 
thin millimeter or submillimeter continuum specific intensity.  Derivation 
of absolute column densities and conversion to volume densities requires 
knowledge or assumptions about dust temperatures, opacities, gas-to-dust 
ratios, geometry, and telescope beam pattern.  Azimuthally averaged (or 
elliptically averaged in cases of extreme axial ratio), 1D intensity profiles 
of starless cores are found to have ``flat" slopes at radii $<$ 3000-7500 
AU and ``steep" slopes at larger radii (see {\it Ward-Thompson et al.}, 1994;
{\it Andr\'e et al.}, 1996; {\it Ward-Thompson et al.}, 1999; {\it Shirley 
et al.}, 2000; {\it Kirk et al.}, 2005).  Such profiles depart from the $n(r) 
\propto r^{-2}$ distribution of the ``singular isothermal sphere" postulated 
by {\it Shu}\ (1977) as the initial state of isolated dense cores prior to 
gravitational collapse.  Similar profiles have been seen also in studies of 
dense core structure utilizing the absorption of background emission by core 
dust (e.g., {\it Alves et al.}, 2001; see chapter by {\it Alves et al.}) 
The observed configurations are better described by ``Bonnor-Ebert 
spheres" (BES; {\it Emden}, 1907; {\it Ebert}, 1955; {\it Bonnor}, 1956;
{\it Chandrasekhar}, 1957), non-singular solutions to the equations of 
hydrostatic equilibrium that can be critically stable in the presence 
of external pressure.  In the 1D isothermal approximation, the family 
of BES solutions are parameterized by the central density, $n_c$, and 
are characterized by density profiles with two regimes: a ``plateau'' 
of slowly decreasing density at small radii and a power-law decrease 
($\sim r^{-2}$) at large radii ({\it Chandrasekhar and Wares}, 1949).   
The size of the plateau scales inversely with the central density, i.e., 
$R(\frac{1}{2}n_c) \propto \sqrt{T/n_c}$.  For reference, Figure 2b 
shows the average density contrast, $\frac{\overline{n}}{n_b}$, where 
$\bar n$ is the average density and $n_b$ is the density at the BES 
(outer) boundary radius, versus BES boundary radius, $R_{b}$, for families 
of BES of given $n_{c}$ ({\it Shirley and Jorgensen}, in preparation).  
Thermally supported density configurations, regardless of geometry, are 
unstable if they have $\frac{\overline{n}}{n_b}$ $\geq 2.46457$ ({\it 
Lombardi and Bertin}, 2001).  Such configurations may remain stable if 
non-thermal internal pressures (e.g., from magnetic fields or turbulence) 
are relevant, however; see Section 5, as well as the next chapter by 
{\it Ward-Thompson et al.}

Detailed modeling of continuum intensity profiles have confirmed that BES 
are a good approximation to the density structures of starless cores (e.g., 
{\it Ward-Thompson et al.}, 1999; {\it Evans et al.}, 2001; {\it Kirk et 
al.}, 2005; {\it Schnee and Goodman}, 2005, but see {\it Harvey et al.}, 
2003).  Recently, {\it Shirley and Jorgensen}\ (in preparation) modeled 
submillimeter continuum emission from 33 isolated starless cores, the 
largest sample so far, and found a log-normal distribution of central 
densities with median $\log n_c$ = 5.3 cm$^{-3}$.  This large median central 
density indicates that non-thermal support (e.g., 
magnetic pressure) has stabilized the cores, that the density scale is 
inaccurate due to the dust opacity assumed (see below), or that these cores 
are not in equilibrium.  Unfortunately, it is not currently possible to 
distinguish between collapsing and static BES with dust emission or 
absorption profiles because the density structure, for a given $n_c$, does 
not significantly vary until late in the collapse history ({\it Myers}, 2005).  
Molecular probes of the kinematics are much better suited to resolving this 
issue (e.g., see Section 5).  Nevertheless, increasing central concentration 
within starless cores could indicate an evolutionary path.  Note, however, 
that while BES are consistent with current observations, this does not prove 
that given physical structures {\it are}\/ BES (cf. {\it Tafalla et al.},
2002; {\it Ballesteros-Paredes et al.}, 2003, but see Section 5.2). 
 
\medskip
\noindent
\textbf{3.3  Temperature Structures}
\medskip

Early calculations of dust temperature ($T_D$) in cores heated only by the 
ISRF indicated $T_D \sim 10$ K ({\it Leung}, 1975), cooler on the inside and 
warmer on the outside.  Such a gradient in $T_D$ is unavoidable theoretically 
because the cores are optically thin to their own radiation 
($\lambda_{\rm{peak}} \approx 200$ $\mu$m) and are primarily heated externally 
by a local ISRF for which short wavelength radiation is significantly 
obnubilated in the center of the core.  
Alternative heating mechanisms for dust, including cosmic rays (i.e., 
direct heating of grains, secondary UV heating from excitation of H$_{2}$, 
and heating of gas followed by gas-grain heating) are not significant 
compared to heating from the ISRF.  More recent calculations using 
current ideas for the ISRF and dust opacites confirm the earlier work 
(e.g., {\it Zucconi et al.}, 2001; {\it Evans et al.}, 2001), finding $T_D$ 
falling from about 12 K on the exposed surface to as low as 7 K at very small 
radii. Such low temperatures would suppress radiation even at millimeter 
wavelengths from the center, possibly affecting interpretation of the central 
densities.  Gradients of $T_D$ found in starless cores, however, do not have 
a strong effect on overall core equilibria, and density profiles very different 
from isothermal BES are not expected ({\it Evans et al.}, 2001; {\it Galli 
et al.}, 2002).

The temperature gradient within a core depends on the strength and spectral 
properties of the ISRF, which are affected by surrounding extinction, and the 
density structure in the core.  Shielded cores have lower temperatures and 
shallower gradients than exposed cores, because the ISRF is attenuated and 
reddened ({\it Stamatellos and Whitworth}, 2003; {\it Andr\'e et al.}, 2003;
{\it Young et al.}, 2004b).  The reddening leaves more of the heating to the 
longer wavelength photons, which ``cook" more evenly.  Cores of relatively 
low central concentration or low central density should have shallower 
temperature gradients due to lower extinction ({\it Shirley et al.}, 2005).  
Calculations in higher dimensions also show smaller gradients.  For clumpy 
cores, the heating is dominated by unobscured lines of sight ({\it Doty et
al.}, 2005b).  Such clumpy cores can have cold patches ($T_D < 6$ K) within 
an overall flatter distribution of $T_D$.

Observational tests do indicate warmer $T_D$ on the outside of cores.  {\it
Langer and Willacy} (2001), {\it Ward-Thompson et al.}\ (2002) and {\it 
Pagani et al.}\ (2004, see also 2003), each using ISO data, found evidence 
for $T_D$ gradients in exposed cores, with innermost $T_{D}$ $\leq$ 10 K.  
Limited spatial resolution at far-infrared wavelengths, however, does not 
allow direct tests of the predictions of very low $T_D$ at very small radii.  
One indirect test comes from studies of starless cores in extinction (e.g., 
{\it Bacmann et al.}, 2000), which are temperature independent.  These studies 
also show ``flat" inner density gradients, and hence these gradients are not 
due to decreased $T_D$ in the centers.  Another possible indirect test is the 
high-resolution study of molecular lines (see below).

Examining gas kinetic temperature ($T_K$), {\it Goldreich and Kwan}\ (1974) 
showed it is forced to be very close to $T_{D}$ at high densities ($n \geq 
10^5$ cm$^{-3}$).  This coupling fails for less dense cores and for the outer 
parts of even denser cores ({\it Goldsmith}, 2001; {\it Galli et al.}, 2002).  
At first, $T_K$ may drop below $T_D$, but cosmic ray heating prevents $T_K$ 
from dropping too low.  If the core is not very deeply shielded, however, 
photoelectric heating causes $T_K$ $>$ $T_{D}$ in the outer layers (e.g., 
{\it Young et al.}, 2004b).  

Reductions in the ISRF affect the chemistry in the outer layers ({\it Lee 
et al.}, 2004b) and change the gas kinetic temperature ($T_K$) profile as 
well by decreasing photoelectric heating.  Conversely, cores in regions 
of high ISRF will have warmer surfaces in both dust and gas ({\it Lis et 
al.}, 2001; {\it J\o rgensen et al.}, 2006).  Since this heating is caused 
by ultraviolet light, it is very sensitive to small amounts of shielding. 
CO is a particularly good constraint on the short-wavelength end of the 
ISRF (e.g., {\it Evans et al.}, 2005), but, with chemical models 
(Section 4) other species can also be excellent probes (e.g., {\it Young et 
al.}, 2004b; {\it Lee et al.}, 2004b).

Observationally, $T_K$ can be determined directly from CO transitions, which 
probe the outer layers, and by inversion transitions of NH$_{3}$, a good 
tracer of dense core material (e.g., see {\it Jijina et al.}, 1999 and 
references therein), which probe deeper layers.  Analyses of such lines 
from starless cores have found nearly constant values of $T_K$ $\approx$ 10 K 
(e.g., {\it Hotzel et al.}, 2002 for B68 and {\it Tafalla et al.}, 2004 for 
L1498 and L1517B). If $T_K = T_D$, and $T_D$ decreases toward the center, 
one might expect to see changes in $T_K$.  For L1498 and L1517B, however, 
{\it Galli et al.}\ (2002) found $T_K$ was $\sim$2 K higher than $T_{D}$ 
for cores with central densities $\sim$10$^{5}$ cm$^{-3}$ (i.e., like L1498 
and L1517B), and that neither $T_K$ nor $T_{D}$ varied significantly within 
the inner core radii.  {\it Shirley et al.}\ (2005), however, found a much 
lower central density (i.e., $n_c = 1-3 \times 10^4$ cm$^{-3}$) for L1498, 
which would explain the difference of $T_K$ and $T_D$.  Also, recent VLA 
observations of NH$_{3}$ (1,1) and (2,2) across the L1544 prestellar core 
by {\it Crapsi et al.} (in preparation) have revealed 
a clear $T_K$ gradient that is consistent with the dust temperature profile 
derived from models of submillimeter continuum emission with a central density 
of 3 $\times$ 10$^{6}$ cm$^{-3}$.

\medskip
\section{\textbf{CHEMICAL CHARACTERISTICS}}
\medskip

Structures of starless cores in principle can be probed using molecular 
emission lines, typically rotational transitions excited at the relatively 
low temperatures and high densities of such objects.  Indeed, line data can 
complement continuum data to provide new constraints to models (e.g., {\it 
Jessop and Ward-Thompson}, 2001).  Line studies also have the advantage of 
tracing kinematic behavior of starless cores.  In practice, however, line 
observations are affected by the abundance variations within the cores and 
by the optical depths of the specific lines used; unlike submillimeter or 
millimeter continuum emission, the optical depths of molecular lines are 
often quite large.  To circumvent this problem, lines from rare isotopologues 
are observed and interpreted using non-LTE radiative transfer codes, which 
take into account the core chemical structure. 
 
Although the basic chemical processes in clouds have been known for some time 
(e.g., formation of H$_{2}$ on dust grains ({\it Gould and Salpeter}, 1963; 
{\it Jura}, 1975), ionization of H$_{2}$ by cosmic rays ({\it Solomon and
Werner}, 1971), ion--molecule reactions ({\it Herbst and Klemperer}, 1973), 
and grain-surface chemistry ({\it Watson and Salpeter}, 1972; {\it Allen and
Robinson}, 1977)), there are still several uncertainties for denser regions, 
one of which is grain-surface processes.  Theoretical studies predicted 
molecular depletion onto grain surfaces (e.g., {\it L\'eger}, 1983).  While 
adsorbed molecules have long been observed in infrared absorption bands (e.g., 
{\it Gillett and Forrest}, 1973; see also {\it Pontoppidan et al.}, 2005), 
only in the past few years have molecular line observations found clear and 
quantitative evidence for gas-phase depletion (e.g., {\it Kuiper et al.}, 
1996; {\it Willacy et al.}, 1998; {\it Kramer et al.}, 1998, see Section 
4.1).  Enhanced deuterium fractionation, which should accompany molecular 
depletion (e.g., {\it Brown and Millar}, 1989), is also found in both 
starless and protostellar cores (see Section 4.2).  Uncertainties still 
remain, however, and further updates of chemical models and detailed 
comparisons with new observational data are needed.  

\medskip
\noindent
\textbf{4.1 Molecular Freeze-out}
\medskip

The dominant gas phase constituent in molecular clouds is H$_{2}$ but its
line emission is not used to trace the interiors of molecular clouds and 
cores.  H$_{2}$, as a homonuclear molecule, has no electric dipole moment 
and thus has no dipole transitions between rotational states.  Quadrupole
emission between states is possible but the upper level of the first allowed 
transition is 512 K above ground (since H$_{2}$ is a light molecule) and 
is only excited where gas is suitably hot, e.g., on cloud surfaces and not
in cloud interiors.

CO and its isotopologues $^{13}$CO, C$^{18}$O or C$^{17}$O are commonly 
used as surrogate tracers to H$_{2}$ since they are relatively abundant and 
intermixed with H$_{2}$, and have rotational transitions excited at the
ambient densities in molecular clouds, i.e., 10$^{2-3}$ cm$^{-3}$.  Recent 
studies, however, have found significant depletions of CO abundance within 
the innermost regions of starless cores by comparing line observations of 
CO isotopologues with the submillimeter dust continuum or extinction maps 
of cores.  For example, {\it Bacmann et al.}\ (2002) found evidence for CO 
depletions by factors of 4-15 in a sample of 7 cores.  Furthermore, {\it 
Caselli et al.}\ (1999) and {\it Bergin et al.}\ (2002), using C$^{17}$O 
and C$^{18}$O respectively, found CO abundance depletions by factors of 
$\sim$1000 and $\sim$100 in the centers of the L1544 and B68 cores 
respectively, i.e., where $n$ $\geq$ 10$^{5}$ cm$^{-3}$.  

The most common explanation for these abundance variations is that CO and 
its isotopologues adsorb, i.e., ``freeze-out," easily onto grain mantles 
at high densities and dust temperatures $<$ 20 K.  Other molecules are 
also similarly depleted onto dust grains within low-mass cores.  For 
example, {\it Bergin et al.}\ (2001), {\it Young et al.}\ (2004b), {\it 
Lai et al.}\ (2003) and {\it Ohashi et al.}\ (1999) found observational 
evidence for depletions of CS, H$_{2}$CO, C$_{3}$H$_{2}$ and CCS respectively 
toward various dense cores.  More recently, {\it Tafalla et al.}\ (in 
preparation) carried out a multimolecular study of the L1498 and L1517B 
cores and found evidence for the depletion of 11 species, including all 
the aforementioned species plus DCO$^+$, HCN, HC$_3$N, HCO$^+$, CH$_3$OH 
and SO. 

Water is also clearly frozen onto dust grains.  First, solid water is seen 
(with abundances $\sim$10$^{-4}$ relative to H$_2$) in absorption within the 
near-infrared spectra of stars located behind molecular clouds and those of 
embedded protostars (e.g., {\it van Dishoeck}, 2004 and references therein).  
Indeed, water ice is the major compound of grain mantles, followed by CO 
(e.g., {\it Chiar et al.}, 1995; {\it Ehrenfreund and Charnley}, 2000).  
Second, the recent SWAS mission furnished stringent upper limits to 
abundances of water vapor ($<$10$^{-8}$ relative to H$_2$; {\it Bergin and
Snell}\ 2002) that are orders of magnitude lower than values predicted by 
gas--phase chemical models that did not account for the gas--grain interaction 
and the consequent molecular freeze-out.  Other ices on grain surfaces 
recently detected through absorption spectra of background stars include 
CO$_{2}$ ({\it Bergin et al.}, 2005), and probably HCOOH and NH$_{4}^{+}$ 
({\it Knez et al.}, 2005).

The degree to which a given molecule depletes onto grains is likely a function 
of its respective binding energy onto grain surfaces, but other chemical 
factors may be also important.  For example, carbon-chain molecules such as 
CCS and C$_3$H$_2$, which are called ``early-time" species because in 
gas--phase chemical models they form before atomic carbon is mainly locked in 
CO ({\it Herbst and Leung}, 1989), can decrease in abundance due to gas-phase 
reactions, especially with atomic oxygen, which tend to convert all C-bearing 
species into CO.  When CO and the reactive O start to freeze-out, carbon-chain 
species increase their abundance again, showing a ``late-time'' secondary 
peak, limited in time by their own adsorption onto grains ({\it Ruffle et 
al.}, 1997, 1999; {\it Li et al.}, 2002; {\it Lee et al.}, 2003).

\begin{figure*}
\epsscale{1.3}
\plotone{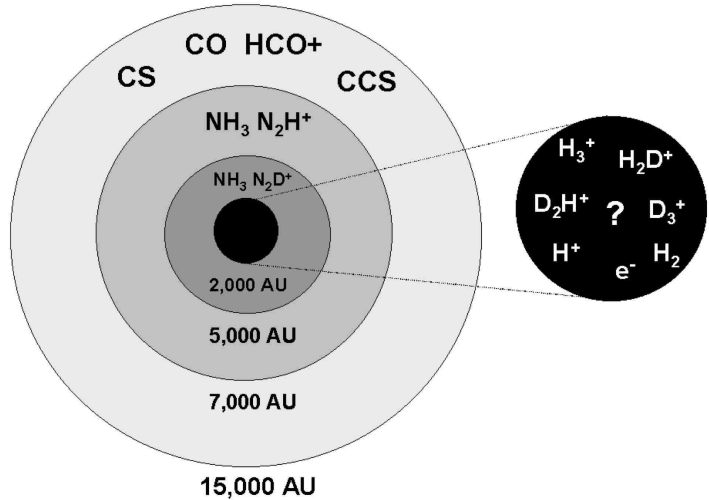}
\caption{\small 
A schematic representation of eventual molecular differentiation within a 
starless core. The external shell of the core (where $n(\rm H_2)$ $\simeq$
10$^4$ cm$^{-3}$) can be traced by CO, CS and other carbon bearing species. 
At radii $<$ 7000~AU, where $n({\rm H_2})$ $\simeq$ 10$^5$ cm$^{-3}$, CO and 
CS disappears from the gas phase and the best gas tracers are NH$_3$ and 
N$_2$H$^+$.  At higher densities, deuterated species become quite abundant 
and, when the density exceeds $\sim$10$^6$ cm$^{-3}$, the chemistry will be 
dominated by light molecular ions, in particular H$_3^+$ and its deuterated 
forms, as well as H$^+$ (e.g., {\it Walmsley et al.}, 2004).} 
\end{figure*}

Nitrogen-bearing molecules, e.g, NH$_3$ and N$_2$H$^+$, have been shown 
to be quite resilient tracers of dense cores, relative to carbon-bearing 
molecules.  They are called ``late-time'' species; they both have N$_2$ as 
a parent molecule, which forms slowly (relative to CO) in the gas phase via 
neutral-neutral reactions.  As described in Section 2, NH$_{3}$ transitions 
have long been tracers of dense cores since these can be excited at densities 
$\approx$ 10$^{4}$ cm$^{-3}$.  In recent years, however, N$_{2}$H$^{+}$ 
transitions have grown in popularity for tracing dense cores with surveys 
by {\it Benson et al.}\ (1998), {\it Caselli et al.}\ (2002c) and {\it 
Tatematsu et al.}\ (2004).  The intensity distribution of N$_{2}$H$^{+}$ 
1--0 in a given core closely matches that of the millimeter or submillimeter 
continuum (e.g., {\it Caselli et al.}, 2002a) and its hyperfine structure 
can be fit to determine excitation temperatures and line opacities (under 
the assumption that all the hyperfine components have the same excitation 
temperature; see {\it Womack et al.}\ 1992; {\it Caselli et al.}\ 1995).  
Within 5 cores, {\it Tafalla et al.}\ (2002; see also {\it Tafalla et al.}, 
2004) found inner zones of CO and CS depletion by factors $\geq$ 100 but 
NH$_{3}$ abundances that increased by factors of $\sim$1-10 towards the 
core centers and constant N$_{2}$H$^{+}$ abundances.  

The relatively high abundances of NH$_{3}$ and N$_{2}$H$^{+}$ in starless 
cores were thought to be due to the low binding energy of their parent N$_{2}$ 
relative to CO ({\it Bergin and Langer}, 1997), but recent laboratory 
experiments by {\it \"Oberg et al.}\ (2005; see also {\it Bisschop et al.}, 
2006) have found the binding energies to be quite similar.  Another possible 
explanation is a slow formation of N$_2$ molecule in the gas phase ({\it 
Caselli and Aikawa}, in preparation).  In the initial formation of molecular 
clouds, the N atom would be the dominant form of nitrogen.  The nitrogen atom 
has a lower adsorption energy than N$_2$ and CO, and thus is more resilient to 
adsorption.  Formation of N$_2$ is much slower than the CO formation in the 
gas phase, and hence many N atoms are still available when the adsorption 
starts (i.e., when the core age exceeds the collisional timescale between gas 
and dust).  Adsorption of N$_{2}$ is compensated by the formation in the gas 
phase.  The high abundance of their mother molecule helps N$_2$H$^+$ and 
NH$_3$ to maintain their abundances.  Two additional factors lead to high 
N$_2$H$^+$ abundance.  First, N$_{2}$H$^+$ does not directly adsorb onto 
grains; it recombines on charged grain surfaces and returns to the gas-phase 
as N$_2$.  Second, the main reactants of N$_{2}$H$^+$, e.g., CO and electrons, 
decrease as the core density rises.  Indeed, in Model A of {\it Aikawa et 
al.}\ (2001), depletion of N$_2$H$^+$ and NH$_3$ is much less significant 
than that of CO, even though the adsorption energies of CO and N$_2$ are 
assumed to be equal.  The central NH$_3$ enhancement relative to N$_2$H$^+$ 
can be caused by CO depletion.  In the outer regions with high CO abundance, 
N$_2$H$^+$ mainly reacts with CO to produce N$_2$, while, in the central 
regions with heavy CO depletion, it recombines to produce NH as well as N$_2$ 
({\it Geppert et al.}, 2004), which is transformed to NH$_3$ ({\it Aikawa et 
al.}, 2005).

Nitrogen-bearing species are not immune to depletion and even N$_{2}$ will 
be eventually adsorbed onto grains at high densities in starless cores (e.g., 
{\it Bergin and Langer}, 1997; {\it Aikawa et al.}, 2001).  {\it Bergin et 
al.}\ (2002) and {\it Crapsi et al.}\ (2004) found evidence for small zones 
of N$_{2}$H$^{+}$ depletion (by factors of $\sim$2) coincident with the 
centers of the exposed cores in B68 and L1521F, respectively.  In addition, 
{\it Di Francesco et al.} (2004) and {\it Pagani et al.}\ (2005) found 
evidence for N$_{2}$H$^{+}$ depletion in the Oph A and L183 cores, 
respectively.  {\it Belloche and Andr\'e}\ (2004) and {\it J\o rgensen et 
al.}\ (2004) found N$_2$H$^+$ depletion within cold IRAM 04191 and NGC 1333
IRAS 2 protostellar cores respectively, but outflows may have also partially 
affected the local chemistry if CO has been released from grain mantles near 
the protostars.  Further, {\it Caselli et al.}\ (2003) and {\it Walmsley et
al.}\ (2004) have suggested that all molecules with C, N or O can be utterly 
depleted in the innermost regions of denser cores based on interpretation of 
the H$_{2}$D$^{+}$ emission in the L1544 core.  Indeed, molecules without 
heavy elements, e.g., H$_{2}$D$^{+}$, may be the only remaining molecular 
tracers of such regions.

Figure 3 schematically summarizes how the eventual molecular differentiation
within a starless core may look.  At radii of $\sim$7000-15000 AU, CO and CS 
are still mainly in the gas phase and the main molecular ion is HCO$^+$, with 
which one can deduce a stringent lower limit of the electron density (e.g., 
Caselli et al., 1998; see Section 4.3).  The fraction of atomic carbon in 
the gas is still large, as testified by the large observed abundances of 
carbon--chain molecules, such as CCS (e.g., {\it Ohashi et al.}, 1999).   
Deeper in the core ($\sim$5000-7000 AU), where the density approaches values 
of the order of 10$^5$ cm$^{-3}$, CO and CS disappear from the gas--phase 
because of the freeze--out onto dust grains.  The physical and chemical 
properties (as well as kinematics) are better traced by N--bearing species, 
in particular NH$_3$.  Within the central 5000~AU, deuterium fractionation 
takes over (see Section 4.2), and N$_2$D$^+$ becomes the best probe ({\it 
Caselli et al.}, 2002a).  NH$_{3}$ is still abundant in these regions, 
however, as suggested by the observed increase of the NH$_3$ abundance 
toward core centers ({\it Tafalla et al.}, 2002).  At $r$ $\leq$ 2500~AU, 
where $n(\rm H_2)$ $\geq$ 10$^6$ cm$^{-3}$, all neutral species are expected 
to freeze--out in short time scales ($\leq$1000 yr) and light species, such 
as H$_3^+$ and its deuterated forms are thought to dominate the chemistry 
and the degree of ionization ({\it Caselli et al.}, 2003; {\it Vastel et 
al.}, 2004). 

The schematic picture of Figure 3 of course depends on the time spent by 
a starless core in this condensed phase, so that one expects to find less 
significant depletion and more typical cloud chemistry in those objects 
that just entered this phase (e.g., possibly the L1521E core; see {\it 
Tafalla and Santiago}, 2004).  Moreover, recent VLA observations of NH$_3$ 
(1,1) and (2,2) by {\it Crapsi et al.}\ (in preparation) have shown that 
NH$_{3}$ is still present in the central 800~AU of the L1544 core, where 
the gas density is a few times 10$^6$ cm$^{-3}$.  If no efficient desorption 
mechanisms are at work, the time spent by the L1544 core nucleus in its high 
density phase may be $<$ 500~yr (see Section 4.4).  More examples are needed 
to refine chemical models of starless cores.

\medskip
\noindent
\textbf{4.2 Deuterium Fractionation}
\medskip

Another important chemical process recently identified within starless cores 
is deuterium fractionation, i.e., the enhancement of deuterated isotopologues 
beyond levels expected from the elemental D/H ratio of $\sim$1.5 $\times$ 
$10^{-5}$ ({\it Oliveira et al.}, 2003).  For example, in a sample of dense 
cores, {\it Bacmann et al.}\ (2003) found a D$_2$CO/H$_2$CO column density 
ratio between 0.01 and 0.1.  In addition, {\it Crapsi et al.}\ (2005) found 
N$_{2}$D$^{+}$/N$_{2}$H$^{+}$ ratios between 0.05 and 0.4 in a similar sample 
of cores.  Deuterium fractionation is related to core temperature and CO 
depletion (e.g., {\it Dalgarno and Lepp}, 1984).  Species such as H$_3^+$ and 
CH$_3^+$ are enriched in deuterium in cold clouds because of the difference in 
zero-point energies between deuterated and non-deuterated species and rapid 
exchange reactions such as H$_3^+$ + HD $\to$ H$_2$D$^+$ + H$_2$ (e.g., {\it 
Millar et al.}, 1989).  The enrichments are propagated to other molecules by 
chemical reactions.  At high densities, heavy element molecules like CO will 
also deplete onto grains, reducing the destruction rate of H$_{2}$D$^{+}$ 
and further increasing the H$_2$D$^+$/H$_3^+$ ratio.  For example, the 
N$_{2}$D$^{+}$/N$_{2}$H$^{+}$ ratio is higher than the DCO$^{+}$/HCO$^{+}$ 
ratio by a factor of about 5 in the L1544 core ({\it Caselli et al.}, 2002b) 
because N$_2$H$^+$ can trace the central region with CO depletion, while 
HCO$^+$, which is produced by the protonation of CO, is not abundant at the 
core center.  

Multiply deuterated species, such as HD$_2^+$ and D$_3^+$, can be similarly
increased in abundance ({\it Roberts et al.}, 2003; see the chapter by {\it 
Ceccarelli et al.}).  For example, the L1544 and IRAS 16293E core centers 
are also rich in H$_2$D$^+$ and HD$_2^+$, as confirmed respectively by {\it 
Caselli et al.}\ (2003) and {\it Vastel et al.} (2004).

The sensitivity of the D/H ratio to physical conditions and molecular 
depletion could make it a potentially useful probe of chemical or dynamic 
evolution of cores.  For example, {\it Crapsi et al.}\ (2005) observed 14 
cores and found that deuterium enhancement was higher in those that are 
more centrally concentrated and have larger peak H$_{2}$ and N$_{2}$H$^{+}$ 
column densities, i.e., arguably those more likely to collapse into stars.  

\medskip
\noindent
\textbf{4.3 Fractional Ionization}
\medskip

Observation of molecular ions, combined with chemical models, are used
frequently to derive ionization degree, $x_{e}$.  This value is important for 
dynamics since the amounts of free charge within cores determine the relative 
influence on core evolution of ambipolar diffusion, the gravitational inward 
motion of neutral species retarded by interactions with ionic species bound to 
a strong magnetic field (e.g., see {\it Ciolek and Mouschovias}, 1995).  For 
example, {\it Caselli et al.}\ (1998; see also {\it Williams et al.}, 1998) 
found, for dense cores observed in DCO$^{+}$ and HCO$^{+}$, $x_{e}$ values 
between 10$^{-8}$ and 10$^{-6}$, with associated ambipolar-diffusion/free-fall 
timescale (AD/FF) ratios of 3-200 and no significant differences between 
starless and protostellar cores.  In the L1544 core, using N$_{2}$D$^{+}$ 
and N$_{2}$H$^{+}$, {\it Caselli et al.}\ (2002b) found a significantly 
smaller $x_{e}$, $\sim$10$^{-9}$, and an associated AD/FF ratio of $\sim$1, 
suggesting this particular core is near dynamical collapse.  

Estimates of $x_{e}$, however, depend on many assumptions, e.g., the 
cosmic-ray ionization rate and molecular depletion degrees, as well as the 
accuracy of the chemical models used.  In particular, the cosmic--ray 
ionization rate, $\zeta$, typically assumed to be around 1--3 $\times$
10$^{-17}$ s$^{-1}$ in molecular clouds (e.g., {\it Herbst and Klemperer}, 
1975; {\it van der Tak and van Dishoeck}, 2000), seems to be closer to 
10$^{-15}$ s$^{-1}$ in diffuse clouds ({\it McCall et al.}, 2003).  
Therefore, a large variation of $\zeta$ is expected in the transition 
between diffuse and molecular clouds (e.g., {\it Padoan and Scalo}, 2005) 
and, maybe, between dense cores with different amounts of shielding.  Such 
variation may, in turn, cause significant variations of the ionization 
degree between cores (e.g., {\it Caselli et al.}, 1998) with consequent 
differences in AD/FF time scales.  {\it Padoan et al.}\ (2004) have 
further argued that the ISRF strengths (i.e., $A_{\rm V}$) and core age 
variations may account for at least some of the large range of $x_{e}$ 
seen in dense cores.  {\it Walmsley et al.}\ (2004) and {\it Flower et
al.}\ (2005) have also suggested that the charge and size distributions 
of dust grains impact ionic abundances, and thus the local AD/FF ratio, 
in the centers of cores depleted of heavy-element species.

\medskip
\noindent
\textbf{4.4 ``Chemodynamical'' Evolution}
\medskip

Depletions or enhancements of various molecular species in starless cores 
can be used as probes of dynamical evolution, since chemical and dynamical 
timescales can significantly differ.  For example, if the dynamical 
timescale of a core is shorter than the adsorption timescale of CO 
($\sim$10$^4 (\frac{10^5 [{\rm cm}^{-3}]}{n({\rm H_2})}) (\frac{10 
[{\rm K}]}{T_{\rm gas}})^{0.5}$ yr), it will not experience CO depletion.  
Recent theoretical studies have investigated the evolution of molecular 
abundance distributions in cores (e.g., in L1544) using models including 
Larson-Penston collapse, ambipolar diffusion of magnetized cores, and 
contraction of a critical BES ({\it Aikawa et al.}, 2001, 2003, 2005; 
{\it Li et al.}, 2002; {\it Shematovich et al.}, 2003; {\it Lee et al.}, 
2004b), showing that chemistry can be used to probe, although not uniquely, 
how the core contracts.  

It is important to note that not all starless cores display similar chemical 
differentiation.  {\it Hirota et al.}\ (2002) found that CCS is centrally 
peaked in the L1521E core, while it is depleted at the center of the L1498 
core ({\it Kuiper et al.}, 1996), although the central density of the L1521E 
core ($\sim$3 $\times$ 10$^{5}$ cm$^{-3}$) is {\it higher} than 
that of the L1498 core ($\sim$1 $\times$ 10$^{4-5}$ cm$^{-3}$; see {\it 
Shirley et al.}, 2005; {\it Shirley and Jorgensen}, in preparation; {\it 
Tafalla et al.}, 2002, 2004, 2005).  The L1521E core also has low NH$_3$ 
and N$_2$H$^+$ abundance, low DNC/HNC ratio, and no CO depletion ({\it Hirota
et al.}, 2001, 2002; {\it Tafalla and Santiago}, 2004).  The core is 
apparently ``chemically young''; these molecular abundances are reproduced 
either at early stages of chemical evolution in a pseudo-time dependent 
model (i.e., assuming that the density does not vary with time), or in a 
dynamical model where the contraction or formation timescale of a core is 
smaller than the chemical timescale (e.g., {\it Lee et al.}, 2003; {\it Aikawa 
et al.}, 2001; 2005).  Other candidates of such ``chemically young'' cores 
are those in L1689B, L1495B, and L1521B ({\it Lee et al.}, 2003; {\it Hirota
et al.}, 2004).  Further determinations of kinematics and physical conditions 
in these cores and ``chemically old" cores (e.g., that in L1498) will be very 
useful toward understanding chemical evolution in starless cores.

\medskip
\section{\textbf{BULK AND TURBULENT MOTIONS}}
\medskip

\noindent
\textbf{5.1  Bulk Motions}
\medskip

\begin{figure*}
\epsscale{1.8}
\plotone{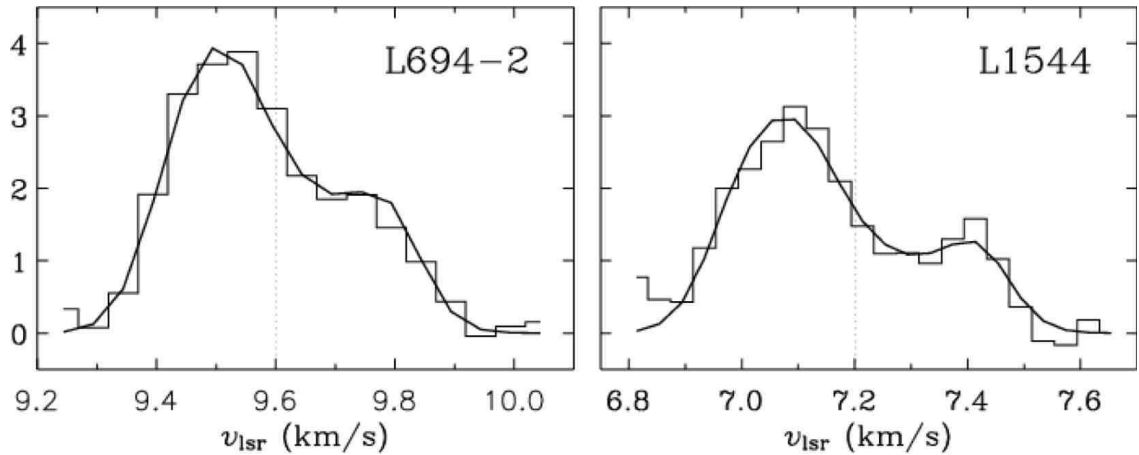}
\caption{\small Spectra of the F1,F = 2,3$\rightarrow$1,2 hyperfine component 
of N$_{2}$H$^{+}$(1--0) (histograms) from the starless cores in L694-2 (left) 
and L1544 (right) from {\it Williams et al.}\ (2006).  Each spectrum is an 
average of emission within the central 10$^{\prime\prime}$ of each core.  
Each profile, asymmetrically blue with respect to the systemic velocity of each 
core (dotted lines), is indicative of inward motions, and can be reproduced 
well with a two-layer radiative transfer model (solid curves).}
\end{figure*}

Molecular line profiles can be effective tracers of starless core kinematics, 
provided appropriate lines are observed and modeled.  For example, internal
velocity gradients in dense cores can be derived from the variation of line 
profiles across cores.  Recent interpretations of such gradients as rotational 
in origin include those made by {\it Barranco and Goodman}\ (1998) of NH$_{3}$ 
(1,1) emission over 3 isolated starless cores, finding velocity gradients of 
$\sim$0.3 to 1.40 km s$^{-1}$ pc$^{-1}$ with projected rotational axes not 
aligned along projected core axes.  In addition, {\it Swift et al.}\ (2005) 
found a 1.2 km s$^{-1}$ pc$^{-1}$ gradient of NH$_{3}$ emission aligned 
with the minor axis of the isolated starless core L1551-MC (see Figure 1).  
Furthermore, {\it Lada et al.}\ (2003) found evidence for 
differential rotation and angular momentum evolution in the B68 core, finding 
velocity gradients of 3.4 km s$^{-1}$ pc$^{-1}$ and 4.8 km s$^{-1}$ pc$^{-1}$ 
from C$^{18}$O (1--0) and N$_{2}$H$^{+}$ (1--0) emission respectively, which 
in turn likely trace outer and inner core material respectively.

Further investigations of velocity gradients include a study of 12 starless 
and 14 protostellar cores mapped in N$_{2}$H$^{+}$ (1--0) by {\it Caselli 
et al.}\ (2002c), who derived a typical velocity gradient of 2 km s$^{-1}$ 
pc$^{-1}$ for both types of cores, and found that the spatial variations of 
the gradients indicate complex motions not consistent with simple solid body 
rotation.  As found in the earlier study of cores by {\it Goodman et al.}\ 
(1993), these motions appear dynamically insignificant, since the derived 
rotational energies are at most a few percent of the respectively derived 
gravitational potential energies.  Even relatively small rotational energies, 
however, could influence protostellar formation during core collapse.  For 
example, angular momentum evolution due to contraction (see below) may 
induce disk formation or fragmentation into multiple objects (e.g., see {\it
Hennebelle et al.}\ 2004).

In addition to plane-of-sky motions, molecular line profiles can be used to 
trace the gas kinematics along the line-of-sight.  For example, {\it Walsh
et al.}\ (2004) determined that the relative motion between 42 isolated cores 
and their surrounding gas was relatively low (e.g., $\le 0.1$ 
km s$^{-1}$) by comparing profiles of a core tracer like N$_{2}$H$^{+}$ with 
profiles of cloud tracers like C$^{18}$O and $^{13}$CO.  This result implies 
that cores do not move ballistically with respect to their surrounding gas.

Molecular line profiles can also be used to detect inward motions in cores.  
Such motions can be detected using optically thick lines, which will appear 
asymmetrically blue relative to symmetric, optically thin lines in a centrally 
contracting system.  Indeed, inward motions may distinguish prestellar cores 
from starless ones, since they may indicate gravitational binding (see Section 
2.1; {\it Gregersen and Evans}, 2000; {\it Crapsi et al.}, 2005).  
Asymmetrically blue profiles were reproduced very well in collapse models of 
the protostellar envelope of B335 by {\it Zhou et al.}\ (1993), {\it Choi et 
al.}\ (1995), and {\it Evans et al.}\ (2005).  In those protostellar cases, 
the models indicated that the self absorption is due to the static envelope 
expected in the inside-out collapse ({\it Shu}, 1977) and the blue peak is 
stronger because of radiative transfer effects (see {\it Zhou et al.}, 1993).  
Although such profiles may in principle also arise from rotational or outflow 
motions, 2D mapping and detailed modeling of the emission can distinguish 
between the different cases (e.g., {\it Narayanan and Walker}, 1998; {\it 
Redman et al.}, 2004).  In addition, explanations other than inward motions 
would not favor blue-skewed over red-skewed profiles in a statistical sample.

Figure 4 shows two examples of such profiles, seen in the F$_1$,F = 2,3 
$\rightarrow$ 1,2 hyperfine component of the N$_{2}$H$^{+}$ (1--0) transition, 
observed towards the central positions of the L694-2 and L1544 starless cores 
from {\it Williams et al.}\ (2006).  Such a blue-skewed profile 
results from red-ward self-absorption, i.e., absorption along the line-of-sight 
by foreground, inward-moving gas of lower excitation.  The statistics and the 
extended nature of many of the blue profiles suggest that inward motions exist 
before formation of the central luminosity source, in contrast to predictions 
of the simplest inside-out collapse models (e.g., {\it Shu}, 1977).  

Such blue-skewed line profiles can be modeled with radiative transfer codes 
to estimate inward motion velocities.  A simplified solution to the problem 
has been presented by {\it Myers et al.}\ (1996) with a ``two-layer" model, 
i.e., with two isothermal gas layers along the line-of-sight approaching each 
other, that can be used to determine rapidly the essentials of inward velocity.
Figure 4 shows how well such modeling can reproduce observed asymmetrically 
blue profiles, with infall velocities of $\sim$0.1 km s$^{-1}$ for both cores.  
With the same goal, {\it De Vries and Myers}\ (2005) have recently provided a 
more realistic ``hill" model, i.e., where the excitation increases with optical 
depth along the line-of-sight.  Full radiative transfer solutions, on the 
other hand, can now be obtained relatively rapidly, and can be used to derive 
velocities (as well as densities, temperatures, and molecular abundances) of 
starless cores (e.g., see {\it Keto et al.}, 2004; {\it Lee et al.}, 2004b).

Surveys of starless cores have revealed many candidates with profiles 
suggestive of inward motions.  For example, {\it Lee et al.}\ (1999), 
using CS (2--1) as an optically thick tracer and N$_{2}$H$^{+}$ 
(1--0) as an optically thin counterpart in single-pointing observations, 
found 17 of 220 optically selected ``starless" cores with asymmetrically 
blue profiles.  Subsequent modeling of these profiles suggested inward 
velocities of 0.04-0.1 km s$^{-1}$, i.e., subsonic.  (The cores here are 
generally isolated and not found in turbulent molecular clouds.)  {\it 
Lee et al.}\ (2004a) also observed 94 cores in the higher excitation CS 
(3--2) and DCO$^{+}$ (2--1) lines, finding 18 strong inward motion 
candidates.  The average inward velocity of 10 cores derived from CS 
(3--2) was found to be slightly larger than the average velocity derived 
from CS (2--1), i.e., 0.07 km s$^{-1}$ vs. 0.04 km s$^{-1}$ respectively, 
suggesting more dense, inner gas moves faster than less dense, outer gas 
in these cores, though the statistical sample was small.  Indeed, CS, being 
depleted in the central and denser regions of the cores (see Section 4), only 
traces the external (radii $\geq$ 7000 AU) envelope of the ``chemically 
old'' cores, so that the extended infall may reflect the (gravity driven) 
motion of the material surrounding these cores.

The nature of the inward motions observed in starless cores can be probed in 
detail through multi-line analyses of individual objects.  For example, the 
isolated core in L1544 was found by {\it Tafalla et al.}\ (1998; see also 
{\it Williams et al.}, 1999) to have CS and H$_{2}$CO line profiles suggesting 
inward motions over a 0.2 pc extent.  This scale is inconsistent with that 
expected from thermal ``inside-out" collapse of a singular isothermal sphere 
since a central protostar of easily detectable luminosity would have likely 
formed in the time required for such a collapse wave to have propagated so 
far.  (A CS/N$_{2}$H$^{+}$ mapping survey of 53 cores by {\it Lee et al.}, 
2001 also found 19 with infall that was very extended.)  In addition, the 
inward velocities determined via two-layer modeling were up to $\sim$0.1 km 
s$^{-1}$.  

Although inward velocities derived from observations are larger than expected 
from some models of ambipolar diffusion, {\it Ciolek and Basu}\ (2000; CB00) 
argued that such motions could indeed result from magnetically retarded 
contraction through a background magnetic field of relatively low strength, 
e.g., $\sim$10 $\mu$G, and tailored a specific ambipolar diffusion model for 
the L1544 core.  Multiple lines of N$_{2}$H$^{+}$ and N$_{2}$D$^{+}$ observed 
from L1544 by {\it Caselli et al.}\ (2002a) indeed showed some consistency with 
features of this model, including the velocity gradient observed across the 
minor axis and a slight broadening of line width and double-peaked line 
profiles at the core centre (possibly due to central depletion, but see {\it 
Myers}, 2005).  Indeed, the ionization fraction of the L1544 core found by 
{\it Caselli et al.}\ (2002b) suggests its AD/FF timescale ratio $\approx$ 
1 (see Section 4.3).  The observed lines at central locations were wider than 
predicted from CB00, however, suggesting possible modifications such as 
turbulence and, in particular, an initial quasi-static layer contraction due 
to Alfv\'enic turbulence dissipation (e.g., {\it Myers and Zweibel}, 2001).  
Still more recent data have revealed further deviations from the CB00 model, 
however.  For example, {\it van der Tak et al.}\ (2005) found evidence in 
double-peaked H$_{2}$D$^{+}$ 1$_{10}$--1$_{11}$ line emission from the L1544 
core for velocities (and densities) that are higher in its innermost regions 
than predicted by the CB00 model.  (The CB00 model could fit N$_{2}$H$^{+}$ 
and H$^{13}$CO$^{+}$ data of the core but they sampled larger radii than 
H$_{2}$D$^{+}$.)  Such velocities could occur if the central region of the 
L1544 core, where thermal pressure dominates, was smaller than predicted by 
CB00 model, or if collapse in these regions has become slightly more dynamic 
(but still subsonic).  

Other isolated cores have shown inward velocities similar to those seen 
in L1544.  For example, {\it Tafalla et al.}\ (2004) found evidence in 
N$_{2}$H$^{+}$ (1--0) line emission toward the L1498 and L1517B cores 
for relatively high velocity gas (i.e., $\sim$0.1 km s$^{-1}$ offset from 
systemic velocities) that was coincident with regions of relatively high 
CS abundance.  Such material may have been accreted onto the core only 
recently, suggesting a more stochastic growth of cores over time.  In 
addition, {\it Williams et al.}\ (2006), comparing N$_{2}$H$^{+}$ 
1--0 of the L694-2 and L1544 cores, found similar inward velocity 
increases with smaller projected radii, although the L694-2 core has a 
shallower gradient but a similar central value to that of the L1544 core 
(see Figure 4).  

Not all starless cores show kinematic behaviour similar to L1544, however.  
For example, {\it Lada et al.}\ (2003) found both blue- and red-skewed 
profiles of CS (2--1) across the B68 core grouped in patterns reminiscent 
of a low-order acoustic oscillation mode.  Since the B68 core may be more 
in structural equilibrium than the L1544 core (e.g., see {\it Alves et al.}, 
2001), simultaneous inward and outward velocities of $\sim$0.1 km s$^{-1}$ 
magnitude in the B68 core may not have to do with core contraction 
and expansion.  Instead, such oscillations may have been excited by external 
influences, e.g., recent passage of a shock from expansion of the Loop I 
superbubble.  

Ambipolar diffusion models like that of CB00 may not be required to explain 
inward motions within starless cores.  For example, {\it Myers}\ (2005) has
recently proposed that the pressure-free, gravitational collapse of 
non-singular, centrally concentrated gas configurations alone is sufficient 
to explain observed velocity features of starless cores; i.e., no magnetic 
retardation is required. 

\medskip
\noindent
\textbf{5.2 Turbulent Motions}
\medskip

Pressure support against gravity within starless cores is provided by both 
thermal and turbulent motions of the gas, in addition to what magnetic field
pressure can impart to neutral species through ions.  Observationally, widths 
of lines from undepleted molecules in starless cores can be used to probe the 
relative influence of thermal and nonthermal (turbulent) motions, and thus 
pressures in such cores, provided gas temperatures can be well determined 
(and magnetic field strengths; see chapter by {\it Ward-Thompson et al.})
Early NH$_{3}$ observations showed that low-mass dense cores have typically 
subsonic levels of gas turbulence ({\it Myers}, 1983).  More recently, {\it 
Goodman et al.}\ (1998; see also {\it Barranco and Goodman}, 1998) found, in 
the cases they examined, that the width of the NH$_{3}$ lines within $\sim$0.1 
pc of the core center does not follow the line width-size relations seen on 
large scales (e.g., $\Delta V \propto R^{0.5}$; see {\it Larson}, 1981), 
suggesting that 
starless cores may constitute localized islands of ``coherence."  {\it 
Goodman et al.}\ further speculated that this reduction of turbulent motions 
may be the result of dense cores at 0.1 pc providing a threshold for turbulent 
dissipation, e.g., through lower ion-neutral coupling from reduced internal 
ionization at high density and extinction.  

Regions of quiescent line width in other starless cores were analyzed by {\it 
Caselli et al.}\ (2002c), who found a lack of general correlation between the 
N$_{2}$H$^{+}$ (1--0) line width and the core impact parameter in a sample 
of 26 cores, in agreement with earlier results.  Furthermore, {\it Tafalla 
et al.}\ (2004; see also {\it Tafalla et al.}, 2002) found a close to constant 
dependence between the NH$_{3}$ or N$_{2}$H$^{+}$ line widths and impact 
parameter in the L1498 and L1517B cores.  This breakdown of the line 
width-size relation has been replicated in recent numerical MHD models of 
dense cores, described elsewhere in this volume; e.g., see chapter by {\it 
Ballesteros-Paredes et al.}  Large numbers of the ``cores" produced in these 
models, although even resembling BES in density configuration, have larger 
internal velocities than the subsonic cores seen in Taurus.  For example, 
{\it Klessen et al.}\ (2005) found only $<$23\% of their model cores were 
subsonic.  Comparisons between observed internal velocities of cores in other 
clouds and those predicted by turbulent models would be very interesting.  
Regions of quiescence within clustered cores have also been noted by {\it 
Williams and Myers}\ (2000) and {\it Di Francesco et al.}\ (2004).  (For 
further discussion of the evolution of turbulent motions in cores, see the 
following chapter by {\it Ward-Thompson et al.})

\medskip
\section{\textbf{CONCLUSIONS}}
\medskip

This review described the numerous recent observations and resulting 
inferences about the nature of starless cores.  These objects represent 
the state of molecular gas preceding gravitational collapse that defines 
the initial conditions of star formation.  Recent observations, together
with recent advances in theory and experimental chemistry, have shown 
that these objects reside at different points on evolutionary paths from 
ambient molecular material to centrally concentrated, slowly contracting 
configurations with significant molecular differentiation.  Though 
significant observational strides have been made, sensitivity and resolution 
limits on past instrumentation have in turn limited our ability to probe 
their internal physical and chemical properties.  Fortunately, new instruments 
such as the wide-field line and continuum focal-plane arrays on single-dish 
telescopes like the JCMT, LMT, and the IRAM 30 m, Nobeyama 45 m, and APEX
Telescopes, and the new sensitive interferometers like the SMA, CARMA, and
ALMA will be available soon.  These will provide the means to characterize 
numerous cores to high degrees, moving our understanding of their internal 
physical and chemical properties from an anecdotal to a statistical regime.

\medskip
\textbf{ Acknowledgments.}  We thank an anonymous referee whose comments
improved this review.  NJE and PCM acknowledge support for this work as 
part of the Spitzer Legacy Science Program, provided by NASA through 
contract 1224608 issued by JPL, California Institute of Technology, under 
NASA contract 1407.  In addition, NJE acknowledges support from NASA OSS 
Grant NNG04GG24G and NSF Grant AST-0307250 to the University of Texas at 
Austin.  PC acknowledges support from the MIUR grant ``Dust and molecules 
in astrophysical environments."  PCM also acknowledges the support of NASA 
OSS Grant NAG5-13050.  
YA acknowledges the support from the COE Program
at Kobe University and Grant-in-Aid for Scientific Research (14740130,
16036205) of MEXT, Japan.

\medskip

\centerline\textbf{REFERENCES}
\medskip
\parskip=0pt
{\small
\baselineskip=11pt


\refs Aikawa Y., Ohashi N., Inutsuka S., Herbst E., and Takakuwa S. 
      (2001) {\it Astrophys.\ J., 552}, 639-653. 

\refs Aikawa Y., Ohashi N., and Herbst E. (2003) {\it Astrophys.\ J., 
      593}, 906-924.

\refs Aikawa Y., Herbst E., Roberts H., and Caselli P. (2005) {\it 
      Astrophys.\ J.}, 620, 330-346. 

\refs Allen L. E., Myers P. C., Di Francesco J., Mathieu R., Chen 
      H., et al. (2002) {\it Astrophys.\ J., 566}, 993-1004.


\refs Allen M. and Robinson G. W. (1977) {\it Astrophys.\ J., 212}, 396-415.

\refs Alves J. F., Lada C. J., and Lada E. A. (2001) {\it Nature, 409}, 
      159-161.

\refs Andr\'e P., Ward-Thompson D. A., and Motte F. (1996) {\it Astron.\ 
      Astrophys., 314}, 625-635.

\refs Andr\'e P., Motte F., and Bacmann A. (1999) {\it Astrophys.\ J., 513}, 
      L57-L60.

\refs Andr\'e P., Ward-Thompson D. A., and Barsony M. (2000) In {\it 
      Protostars and Planets IV}\ (V. Mannings et al., eds.), pp. 59-96.  
      Univ. of Arizona, Tucson.

\refs Andr\'e P., Bouwman J., Belloche A., and Hennebelle P. (2003) In 
      {\it Chemistry as a Diagnostic of Star Formation}\ (C. L. Curry and
      M. Fich, eds.), pp. 127-138.  NRC Press, Ottawa.


\refs Bacmann A., Andr\'e P., Puget J.-L., Abergel A., Bontemps S., 
      et al. (2000) {\it Astron.\ Astrophys., 361}, 555-580.


\refs Bacmann A., Lefloch B., Ceccarelli C., Castets A., Steinacker J.,
      et al. (2002) {\it Astron.\ Astrophys., 389}, L6-L10.


\refs Bacmann A., Lefloch B., Ceccarelli C., Steinacker J., Castets 
      A., et al. (2003) {\it Astrophys.\ J., 585}, L55-L58.


\refs Ballesteros-Paredes J., Klessen R.~S., and V{\'a}zquez-Semadeni E.\ 
      (2003) {\it Astrophys.\ J., 592}, 188-202.

\refs Barranco J. A. and Goodman A. A. (1998) {\it Astrophys.\ J., 504}, 
      207-222.

\refs Beichman C.~A., Myers P.~C., Emerson J.~P., Harris S., Mathieu R., 
      et al. (1986) {\it Astrophys.\ J., 307}, 337-339.


\refs Belloche A. and Andr\'e P. (2004) {\it Astron.\ Astrophys., 419}, 
      L35-L38.


\refs Benson P. J. and Myers P. C. (1989) {\it Astrophys.\ J.\ Suppl., 71}, 
      89-108.

\refs Benson P. J., Caselli P., and Myers P. C. (1998) {\it Astrophys.\ J.,
      506}, 743-757.

\refs Bergin E. A. and Langer W. D. (1997) {\it Astrophys.\ J., 486}, 
      316-328.

\refs Bergin E. A. and Snell R. L. (2002) {\it Astrophys.\ J., 581}, 
      L105-L108.

\refs Bergin E. A., Plume R., Williams J. P., and Myers P. C. (1999)
      {\it Astrophys.\ J., 512}, 724-739.

\refs Bergin E. A., Ciardi D. R., Lada C. J., Alves J., and Lada E. A. 
      (2001) {\it Astrophys.\ J., 557}, 209-225.

\refs Bergin E. A., Alves J., Huard T., and Lada C. J. (2002) {\it 
      Astrophys.\ J., 570}, L101-L104.

\refs Bergin E. A., Melnick G. J., Gerakines P. A., Neufeld D. A., and
      Whittet D. C. B. (2005) {\it Astrophys.\ J., 627}, L33-L36.


\refs Bisschop S. E., Fraser H. J., \"Oberg K. I., van Dishoeck E. F.,
      and Schlemmer S. (2006) {\it Astron.\ Astrophys.}, in press. 

\refs Black J. H. (1994) In {\it ASP Conf. Ser. 58, The First Symposium on 
      the Infrared Cirrus and Diffuse Interstellar Clouds}\ (R. M. Cutri and 
      W. B. Latter, eds.), pp. 355-367. ASP, San Francisco.

\refs Bonnor W. B. (1956) {\it Mon.\ Not.\ R.\ Astron.\ Soc., 116}, 351-359.





\refs Brown P. D. and Millar T. J. (1989) {\it Mon.\ Not.\ R.\ Astron.\ Soc., 
      240}, 25-29.


\refs Caselli P., Myers P. C., and Thaddeus P. (1995) {\it Astrophys.\ J.,
      455}, L77-L80.

\refs Caselli P., Walmsley C. M., Terzieva R., and Herbst E. (1998)
      {\it Astrophys.\ J., 499}, 234-249. 

\refs Caselli P., Walmsley C. M., Tafalla M., Dore L., and Myers 
      P. C. (1999) {\it Astrophys.\ J., 523}, L165-L169. 

\refs Caselli P., Walmsley C. M., Zucconi A., Tafalla M., Dore L., et al.
      (2002a) {\it Astrophys.\ J., 565}, 331-343.


\refs Caselli P., Walmsley C. M., Zucconi A., Tafalla M., Dore L., et al. 
      (2002b) {\it Astrophys.\ J., 565}, 344-358.


\refs Caselli P., Benson P. J., Myers P. C., and Tafalla M. (2002c) {\it 
      Astrophys.\ J., 572}, 238-263.

\refs Caselli P., van der Tak F. F. S., Ceccarelli C., and Bacmann
      A. (2003) {\it Astron.\ Astrophys., 403}, L37-L41. 


\refs Chandrasekhar S. (1957) ``An Introduction to the Study of Stellar 
      Structure", (Dover: Chicago), p. 155

\refs Chandrasekhar S. and Wares G. W. (1949) {\it Astrophys.\ J., 109}, 
      551-554. 

\refs Chiar J. E., Adamson A. J., Kerr T. H., and Whittet D. C. B.
      (1995) {\it Astrophys.\ J., 455}, 234-243.

\refs Chini R., Reipurth B., Ward-Thompson D., Bally J., Nyman L.-\AA., 
      et al. (1997) {\it Astrophys.\ J., 474}, L135-L138.


\refs Chini R., K\"ampgen K., Reipurth B., Albrecht M., Kreysa E.,
      et al. (2003) {\it Astron.\ Astrophys., 409}, 235-244.


\refs Choi M., Evans N. J., II, Gregersen E. M., and Wang Y. (1995) 
      {\it Astrophys.\ J., 448,} 742-747.

\refs Ciolek G. E. and Basu S. (2000) {\it Astrophys.\ J., 529}, 925-931.

\refs Ciolek G. E. and Mouschovias T. Ch. (1995) {\it Astrophys.\ J., 
      454}, 194-216.

\refs Clemens D.~P. and Barvainis R.\ (1988) {\it Astrophys.\ J.\ Suppl., 
      68}, 257-286. 

\refs Crapsi A., Caselli P., Walmsley C. M., Myers P. C., Tafalla M., 
      et al.  (2005) {\it Astrophys.\ J., 619}, 379-406.





\refs Dalgarno A. and Lepp S. (1984) {\it Astrophys.\ J., 287}, L47-L50.

\refs De Vries C. H. and Myers P. C. (2005) {\it Astrophys.\ J., 620}, 
      800-815. 



\refs Di Francesco J., Myers P. C., and Andr\'e P. (2004) {\it Astrophys.\ 
      J., 617}, 425-438.


\refs Doty S. D., Everett S. E., Shirley Y. L., Evans N. J., II, and
      Palotti M. L. (2005a) {\it Mon.\ Not.\ R.\ Astron.\ Soc., 359}, 
      228-236.

\refs Doty S. D., Metzler R. A., and Palotti M. L. (2005b) {\it Mon.\ Not.\ 
      R.\ Astron.\ Soc., 362}, 737-747.

\refs Draine B. T. (1978) {\it Astrophys.\ J.\ Suppl., 36}, 595-619.

\refs Draine B. T. and Lee H. M. (1984) {\it Astrophys.\ J., 285}, 89-108.


\refs Ebert R. (1955) ZAp, 37, 217-232.

\refs Egan M. P., Leung C. M., and Spagna G. F. (1988) {\it Comp.\ Phys.\
      Comm., 48}, 271-292.

\refs Ehrenfreund P. and Charnley S. B. (2000) {\it Ann.\ Rev.\ Astron.\
      Astrophys., 38}, 427-483.

\refs Emden R. (1907) {\it Gaskugeln}, Leipzig and Berlin, Table 14.

\refs Enoch M. L., Young K. E., Glenn J., Evans N. J., II, Golwala S., et al.
      (2006) {\it Astrophys.\ J.}, in press.

\refs Evans N. J., II, Rawlings J. M. C., Shirley Y. L., and Mundy L. G.
      (2001) {\it Astrophys. J., 557}, 193-208.

\refs Evans N. J., II, Lee J.-E., Rawlings J. M. C., and Choi M. (2005)
      {\it Astrophys.\ J., 626}, 919-932.

\refs Flower D. R., Pineau des For\^ets G., and Walmsley C. M. (2005)
      {\it Astron. Astrophys., 436}, 933-943.

\refs Galli D., Walmsley M., and Gon\c calves J. (2002) {\it Astron.\ 
      Astrophys., 394}, 275-284.

\refs Geppert W. D., Thomas R., Semaniak J., Ehlerding A., Millar T. J., 
      et al. (2004) {\it Astrophys.\ J., 609}, 459-464.
       
       
\refs Gillett F. C. and Forrest W. J. (1973) {\it Astrophys.\ J., 179}, 483

\refs Goldreich P. and Kwan J.\ (1974) {\it Astrophys.\ J., 189}, 441-453.

\refs Goldsmith P. F. (2001) {\it Astrophys.\ J., 557}, 736-746.

\refs Goodman A. A., Benson P. J., Fuller G. A., and Myers P. C. (1993)
      {\it Astrophys.\ J., 406}, 528-547.

\refs Goodman A. A., Barranco J. A., Wilner D. J., and Heyer M. H.
      (1998) {\it Astrophys.\ J., 504}, 223-246.

\refs Goodwin S. P., Ward-Thompson D., and Whitworth A. P. (2002) {\it 
      Mon.\ Not.\ R.\ Astron.\ Soc., 330}, 769-771.

\refs Gould R. J. and Salpeter E. E. (1963) {\it Astrophys.\ J., 138}, 
      393-407

\refs Gregersen E.~M. and Evans N.~J.\ (2000) {\it Astrophys.\ J., 538}, 
      260-267. 


\refs Hara A., Tachihara K., Mizuno A., Onishi T., Kawamura A., 
      et al. (1999) {\it Pub.\ Astron.\ Soc.\ Japan, 51}, 895-910.


\refs Hartmann L. (2002) {\it Astrophys.\ J., 578}, 914-924.

\refs Harvey D. W. A., Wilner D. J., Di Francesco J., Lee C. W., 
      Myers P. C., et al. (2002) {\it Astron.\ J., 123}, 3325-3328.



\refs Harvey D. W. A., Wilner D. J., Myers P. C., and Tafalla M. 
      (2003) {\it Astrophys.\ J., 597}, 424-433.

\refs Hatchell J., Richer J. S., Fuller G. A., Qualtrough C. J., Ladd 
      E. F., et al. (2005) {\it Astron.\ Astrophys., 440}, 151-161.



\refs Hennebelle P., Whitworth A. P., Cha S.-H., and Goodwin S. P. 
      (2004) {\it Mon.\ Not.\ R.\ Astron.\ Soc., 348}, 687-701.

\refs Herbst E. and Klemperer W. (1973) {\it Astrophys.\ J., 185}, 505-534.

\refs Herbst E. and Leung C. M. (1989) {\it Astrophys.\ J.\ Suppl., 69}, 
      271-300.

\refs Hirota T., Ikeda M., and Yamamoto S. (2001) {\it Astrophys.\ J., 547}, 
      814-828.

\refs Hirota T., Ito T., and Yamamoto S. (2002) {\it Astrophys.\ J., 565}, 
      359-363.

\refs Hirota T., Maezawa H., and Yamamoto S. (2004) {\it Astrophys.\ J., 
      617}, 399-405.


\refs Hotzel S., Harju J., and Juvela M. (2002) {\it Astron.\ Astrophys., 
      395}, L5-L8.




\refs Ivezi\'{c} \u{Z}., Nenkova M., and Elitzur M. (1999) {\it User Manual 
      for DUSTY}, Univ. of Kentucky Internal Report.

\refs Jessop N. E. and Ward-Thompson D. (2000) {\it Mon.\ Not.\ R.\ Astron.\ 
      Soc., 311}, 63-74.

\refs Jessop N. E. and Ward-Thompson D. (2001) {\it Mon.\ Not.\ R.\ Astron.\ 
      Soc., 323}, 1025-1034.

\refs Jijina J., Myers P. C., and Adams F. C. (1999) {\it Astrophys.\ J.\ 
      Suppl., 125}, 161-236.

\refs Johnstone D. and Bally J. (1999) {\it Astrophys. J., 510}, L49-L53.

\refs Johnstone D., Wilson C. D., Moriarty-Schieven G., Joncas
      G., Smith G., et al. (2000) {\it Astrophys.\ J., 545}, 327-339.


\refs Johnstone D., Fich M., Mitchell G. F., and Moriarty-Schieven G. (2001) 
      {\it Astrophys.\ J., 559}, 307-317.

\refs Johnstone D., Di Francesco J., and Kirk H. M. (2004) {\it Astrophys.\ 
      J., 611}, L45-L48.

\refs Jones C. E., Basu S., and Dubinski J. (2001) {\it Astrophys.\ J., 
      551}, 387-393.

\refs J\o rgensen J. K., Hogerheijde M. R., van Dishoeck E. F., Blake 
      G. A., and Sch\"oier F. L. (2004) {\it Astron.\ Astrophys., 413}, 
      993-1007.

\refs J\o rgensen J. K., Johnstone D., van Dishoeck E. F., and Doty S. D.
      (2006) {\it Astron.\ Astrophys.}, in press.

\refs Jura M. (1974) {\it Astrophys.\ J., 191}, 375-379.

\refs Kato S., Mizuno N., Asayama S., Mizuno A., Ogawa H., et al. (1999) 
      {\it Pub.\ Astron.\ Soc.\ Japan, 51}, 883-893.


\refs Kauffmann J., Bertoldi F., Evans N. J., II, et al. (2005) 
      {\it Astron.\ Nach., 326}, 878-881.
   
   
\refs Keto E., Rybicki G. B., Bergin E. A., and Plume R. (2004) {\it
      Astrophys.\ J., 613}, 355-373.

\refs Kirk H. M., Johnstone D., and Di Francesco J. (2006), in press.

\refs Kirk J. M., Ward-Thompson D., and Andr\'e P. (2005) {\it Mon.\ Not.\ 
      R.\ Astron.\ Soc., 360}, 1506-1526.

\refs Klessen R. S., Ballesteros-Paredes J., V\'azquez-Semadeni E., and 
      Dur\'an-Rojas C. (2005) {\it Astrophys.\ J.}, 620, 786-794.

\refs Knez C., Boogert A. C. A., Pontoppidan K. M., Kessler-Silacci J. E.,
      van Dishoeck E. F., et al. (2005) {\it Astrophys.\ J., 635}, L145-L148.


\refs Kramer C., Alves J., Lada C. J., Lada E. A., Sievers A., 
      et al. (1998) {\it Astron.\ Astrophys., 329}, L33-L36.


\refs Kramer C., Richer J., Mookerjea B., Alves J., and Lada C. (2003), 
      {\it Astron.\ Astrophys., 399}, 1073-1082.

\refs Kuiper T. B. H., Langer W. D., and Velusamy T. (1996) {\it Astrophys.\
      J., 468}, 761-773.


\refs Lada C. J. and Lada E. A. (2003) {\it Ann.\ Rev.\ Astron.\ Astrophys., 
      41}, 57-115.

\refs Lada C. J., Bergin E. A., Alves J. F., and Huard T. L. (2003) {\it 
      Astrophys.\ J., 586}, 286-295.



\refs Lai S.-P., Velusamy T., Langer W. D., and Kuiper T. B. H. (2003) {\it 
      Astron.\ J., 126}, 311-318.

\refs Larson R. B. (1981) {\it Mon.\ Not.\ R.\ Astron.\ Soc., 194}, 809-826.

\refs Langer W. D. and Willacy K. (2001) {\it Astrophys.\ J., 557}, 714-726.

\refs Langer W. D., van Dishoeck E. F., Bergin E. A., Blake G. A., Tielens A. 
      G. G. M., et al. (2000) In {\it Protostars and Planets IV}\ (V. Mannings
      et al., eds.), pp. 29-57. Univ. of Arizona, Tucson.

\refs Lee C. W. and Myers P. C. (1999) {\it Astrophys.\ J.\ Suppl., 123}, 
      233-250.

\refs Lee C. W., Myers P. C., and Tafalla M. (1999) {\it Astrophys.\ J., 
      526}, 788-805.

\refs Lee C. W., Myers P. C., and Tafalla M. (2001) {\it Astrophys.\ J.\ 
      Suppl., 136}, 703-734.

\refs Lee C. W., Myers P. C., and Plume R. (2004a) {\it Astrophys.\ J.\ 
      Suppl., 153}, 523-543. 

\refs Lee J.-E., Evans N. J., II, Shirley Y. L., and Tatematsu K. (2003) 
      {\it Astrophys.\ J., 583}, 789-808.

\refs Lee J.-E., Bergin E. A., and Evans N. J., II (2004b) {\it Astrophys.\ 
      J., 617}, 360-383.


\refs L\'eger A. (1983) {\it Astron.\ Astrophys., 123}, 271-278.

\refs Leung C.~M.\ (1975) {\it Astrophys.\ J., 199}, 340-360.

\refs Li Z.-Y. and Shu F. H. (1996) {\it Astrophys.\ J., 472}, 211-224.

\refs Li Z.-Y., Shematovich V. I., Wiebe D. S., and Shustov B. M. (2002)
      {\it Astrophys.\ J., 569}, 792-802.
  
\refs Lis D. C., Serabyn E., Keene J., Dowell C. D., Benford D. J., 
      et al. (1998) {\it Astrophys.\ J., 509}, 299-308.


\refs Lis D. C., Serabyn E., Zylka R., and Li Y.\ (2001) {\it Astrophys.\ 
      J., 550}, 761-777.

\refs Lombardi M. and Bertin G. (2001) {\it Astron.\ Astrophys., 375}, 
      1091-1099.

\refs Lynds B.~T.\ (1962) {\it Astrophys.\ J.\ Suppl., 7}, 1-52.


\refs Mathis J. S., Mezger P. G., and Panagia N. (1983) {\it Astron.\ 
      Astrophys., 128}, 212-229.

\refs McCall B. J., Huneycutt A. J., Saykally R. J., Geballe T. R., Djuric N.,
      et al. (2003) {\it Nature, 422}, 500-502.

\refs Millar T. J., Bennett A., and Herbst E. (1989) {\it Astrophys.\ 
      J., 340}, 906-920.


\refs Mitchell G. F., Johnstone D., Moriarty-Schieven G., Fich M.,
      and Tothill N. F. H. (2001) {\it Astrophys.\ J., 556}, 215-229.

\refs Mizuno A., Hayakawa T., Tachihara K., Onishi T., Yonekura Y., 
      et al. (1999) {\it Pub.\ Astron.\ Soc.\ Japan, 51}, 859-870.



\refs Motte F., Andr\'e P., and Neri R.\ (1998) {\it Astron.\ Astrophys., 
      336}, 150-172.

\refs Motte F., Andr\'e P., Ward-Thompson D., and Bontemps S. (2001)
      {\it Astron.\ Astrophys., 372}, L41-L44.

\refs Murphy D. C. and Myers P. C. (2003) {\it Astrophys.\ J., 591}, 
      1034-1048.

\refs Myers P. C. (1983) {\it Astrophys.\ J., 270}, 105-118.



\refs Myers P. C. (2005) {\it Astrophys.\ J., 623}, 280-290.

\refs Myers P. C. and Benson P. J. (1983) {\it Astrophys.\ J., 266}, 
      309-320.

\refs Myers P. C. and Zweibel E. (2001) {\it Bull.\ Am.\ Astron.\ Soc., 
      33}, 915.

\refs Myers P.~C., Linke R.~A., and Benson P.~J.\ (1983) {\it Astrophys.\ 
      J., 264}, 517-537.

\refs Myers P. C., Fuller G. A., Goodman A. A., and Benson P. J. 
      (1991) {\it Astrophys.\ J., 376}, 561-572.

\refs Myers P. C., Mardones D., Tafalla M., Williams J. P., and Wilner 
      D. J. (1996) {\it Astrophys.\ J., 465}, L133-L136.

\refs Narayanan G. and Walker C. K. (1998) {\it Astrophys.\ J., 508}, 
      780-790.


\refs Nutter D. J., Ward-Thompson D., and Andr\'e P. (2005) {\it Mon.\ 
      Not.\ R.\ Astron.\ Soc., 357}, 975-982.

\refs Obayashi A., Kun M., Sato F., Yonekura Y., and Fukui Y. (1998) 
      {\it Astron.\ J., 115}, 274-285.

\refs \"Oberg K. I., van Broekhuizen F., Fraser H. J., Bisschop S. E.,
      van Dishoeck E. F., et al. (2005) {\it Astrophys.\ J., 621}, L33-L36. 


\refs Ohashi N., Lee S. W., Wilner D. J., and Hayashi M. (1999) {\it 
      Astrophys.\ J., 518}, L41-L44. 

\refs Oliveira C. M., H\'ebrard G., Howk C. J., Kruk J. W., Chayer P., et 
      al. (2003) {\it Astrophys.\ J., 587}, 235-255.

\refs Onishi T., Kawamura A., Abe R., Yamaguchi N., Saito H., 
      et al. (1999) {\it Pub.\ Astron.\ Soc.\ Japan, 51}, 871-881.


\refs Onishi T., Mizuno A., Kawamura A., Ogawa H., and Fukui Y. (1998)
      {\it Astrophys.\ J., 502}, 296-314.

\refs Onishi T., Mizuno A., Kawamura A., Tachihara K., and Fukui 
      Y. (2002) {\it Astrophys.\ J., 575}, 950-973.

\refs Ossenkopf V. and Henning T. (1994) {\it Astron.\ Astrophys., 291}, 
      943-959.

\refs Padoan P. and Scalo J. (2005) {\it Astrophys.\ J., 624}, L97-L100.

\refs Padoan P., Willacy K., Langer W. D., and Juvela M. (2004) {\it 
      Astrophys.\ J., 614}, 203-210.

\refs Pagani L., Lagache G., Bacmann A., Motte F., Cambr\'esy L., et al.
      (2003) {\it Astron.\ Astrophys., 406}, L59-L62.


\refs Pagani L., Bacmann A., Motte F., Cambr\'esy L., Fich M., et al.
      (2004) {\it Astron.\ Astrophys., 417}, 605-613.


\refs Pagani L., Pardo J.-R., Apponi A. J., Bacmann A., and Cabrit S.
      (2005) {\it Astron.\ Astrophys., 429}, 181-192.




\refs Pollack J. B., Hollenbach D., Beckwith S., Simonelli D. P., Roush T.,
      et al. (1994) {\it Astrophys.\ J., 421}, 615-639.

\refs Pontoppidan K. M., Dullemond C. P., van Dishoeck E. F., Blake 
      G. A., Boogert A. C. A., Evans N, J., II, et al. (2005) {\it 
      Astrophys.\ J., 622}, 463-481.


\refs Redman M. P., Keto E., Rawlings J. M. C., and Williams D. A. 
      (2004) {\it Mon.\ Not.\ R.\ Astron.\ Soc., 352}, 1365-1371.


\refs Roberts H., Herbst E., and Millar T. J. (2003) {\it Astrophys.\ 
      J., 591}, L41-L44.

\refs Ruffle D. P., Hartquist T. W., Taylor S. D., and Williams D. A. 
      (1997) {\it Mon.\ Not.\ R.\ Astron.\ Soc., 291}, 235-240.

\refs Ruffle D. P., Hartquist T. W., Caselli P., and Williams D. A. 
      (1999) {\it Mon.\ Not.\ R.\ Astron.\ Soc., 306}, 691-695.

\refs Ryden B. S. (1996) {\it Astrophys.\ J., 471}, 822-831.



\refs Sandell G. and Knee L. B. G. (2001) {\it Astrophys.\ J., 546}, 
      L49-L52.

\refs Schnee S. and Goodman A. A. (2005) {\it Astrophys.\ J., 624}, 254-266.

\refs Shematovich V. I., Wiebe D. S., Shustov B. M., and Li Z.-Y. 
      (2003) {\it Astrophys.\ J., 588}, 894-909. 


\refs Shirley Y. L., Evans N. J., II, Rawlings J. M. C., and Gregersen 
      E. M. (2000) {\it Astrophys.\ J.\ Suppl., 131}, 249-271.

\refs Shirley Y. L., Nordhaus M. K., Grcevich J. M., Evans N. J., II, 
      Rawlings J. M. C., et al. (2005) {\it Astrophys.\ J., 632}, 982-1000.



\refs Shu F. H. (1977) {\it Astrophys.\ J., 214}, 488-497.


\refs Solomon P. M. and Werner M. W. (1971) {\it Astrophys.\ J., 165},
      41-49.

\refs Stamatellos D. and Whitworth A. P. (2003) {\it Astron.\ Astrophys., 
      407}, 941-955.

\refs Stamatellos D., Whitworth A. P., Andr\'e P., and Ward-Thompson 
      D. (2004) {\it Astron.\ Astrophys., 420}, 1009-1023.


\refs Stanke T., Smith M. D., Gredel R., and Khanzadyan T. (2006) 
      {\it Astron.\ Astrophys.}, in press. 


\refs Steinacker J., Bacmann A., Henning Th., Klessen R., and Stickel 
      M. (2005) {\it Astron.\ Astrophys., 434}, 167-180.

\refs Swift J., Welch W. J., and Di Francesco J. (2005a) {\it Astrophys.\ 
      J., 620}, 823-834.


\refs Tachihara K., Mizuno A., and Fukui Y. (2000) {\it Astrophys.\ J., 
      528}, 817-840.

\refs Tachihara K., Onishi T., Mizuno A., and Fukui Y. (2002) {\it 
      Astron.\ Astrophys., 385}, 909-920.

\refs Tafalla M. and Santiago J. (2004) {\it Astron.\ Astrophys., 414}, 
      L53-L56.

\refs Tafalla M., Mardones D., Myers P. C., Caselli P., Bachiller R.,
      et al. (1998) {\it Astrophys.\ J., 504}, 900-914.


\refs Tafalla M., Myers P. C., Caselli P., Walmsley C. M., and Comito C. 
      (2002) {\it Astrophys.\ J., 569}, 815-835.

\refs Tafalla M., Myers P. C., Caselli P., and Walmsley C. M. (2004) 
      {\it Astron.\ Astrophys., 416}, 191-212.



\refs Tatematsu K., Umemoto T., Kandori R., and Sekimoto Y. (2004) 
      {\it Astrophys.\ J., 606}, 333-340.




\refs T\'oth L. V., Haas M., Lemke D., Mattila K., and Onishi T. (2004)
      {\it Astron.\ Astrophys., 420}, 533-546.

\refs van der Tak F. F. S. and van Dishoeck E. F. (2000) {\it Astron.\ 
      Astrophys., 358}, L79-L82.

\refs van der Tak F. F. S., Caselli P., and Ceccarelli C. (2005) 
      {\it Astron.\ Astrophys., 439}, 195-203.

\refs van Dishoeck E. F. (1988) In {\it Rate Coefficients in Astrochemistry}\ 
      (T. L. Millar and D. A. Williams, eds.), p. 49-62. Kluwer, Dordrecht.

\refs van Dishoeck E. F. (2004) {\it Ann.\ Rev.\ Astron.\ Astrophys., 
      42}, 119-167.

\refs Vastel C., Phillips T. G., and Yoshida H. (2004) {\it Astrophys.\ 
      J., 606}, L127-L130.

\refs Walmsley C. M., Flower D. R., and Pineau des For\^ets G. (2004) 
      {\it Astron.\ Astrophys., 418}, 1035-1043.

\refs Walsh A. J., Myers P. C., and Burton M. G. (2004) {\it Astrophys.\ 
      J., 614}, 194-202.

\refs Ward-Thompson D., Scott P. F., Hills R. E., and Andr\'e P. (1994)
      {\it Mon.\ Not.\ R.\ Astron.\ Soc., 268}, 276-290.

\refs Ward-Thompson D., Motte F., and Andr\'e P. (1999) {\it Mon.\ Not.\ 
      R.\ Astron.\ Soc., 305}, 143-150.


\refs Ward-Thompson D., Andr\'e P., and Kirk J. M. (2002) {\it Mon.\ 
      Not.\ R.\ Astron.\ Soc., 329}, 257-276.

\refs Watson W. D. and Salpeter E. E. (1972) {\it Astrophys.\ J., 174}, 
      321-340.




\refs Williams J. P. and Myers P. C. (1999) {\it Astrophys.\ J., 518},
      L37-L40.

\refs Williams J. P., de Geus E. J., and Blitz L. (1994) {\it Astrophys.\ 
      J., 428}, 693-712.

\refs Williams J. P., Bergin E. A., Caselli P., Myers P. C., and
      Plume R. (1998) {\it Astrophys. J., 503}, 689-699.

\refs Williams J. P., Myers P. C., Wilner D. J., and Di Francesco J. 
      (1999) {\it Astrophys.\ J., 513}, L61-L64.

\refs Williams J. P., Lee C. W., and Myers P. C. (2006) {\it Astrophys.\ 
      J., 636}, 952-958.

\refs Willacy K., Langer W. D., and Velusamy T. (1998) {\it Astrophys.\ 
      J., 507}, L171-L175.

\refs Wilson C. D., Avery L. W., Fich M., Johnstone D., Joncas G., et al.
      (1999) {\it Astrophys.\ J., 513}, L139-L142.


\refs Womack M., Ziurys L. M., and Wyckoff S. (1992) {\it Astrophys.\ J., 
      387}, 417-429.

\refs Yonekura Y., Mizuno N., Saito H., Mizuno A., Ogawa H., et al. (1999) 
      {\it Pub.\ Astron.\ Soc.\ Japan, 51}, 911-918.

\refs Young C.~H., J\o rgensen J. K., Shirley Y. L., Kauffmann J., Huard T.,
      et al. (2004a) {\it Astrophys.\ J.\ Suppl., 154}, 396-401.

\refs Young K.~E., Lee J.-E., Evans N.~J., Goldsmith P.~F., and Doty 
      S.~D.\ (2004b) {\it Astrophys.\ J., 614}, 252-266. 

\refs Young K. E., Enoch M. L., Evans N. J., II, Glenn J., Sargent A. I.,
      et al. (2006) {\it Astrophys.\ J.}, in press.

\refs Yun J.~L. and Clemens D.~P.\ (1990) {\it Astrophys.\ J., 365}, L73-L76.

\refs Yun J.~L., Moreira M. C., Torrelles J. M., Afonso J. M., and 
      Santos N. C. (1996) {\it Astron.\ J., 111}, 841-845.

\refs Zhou S., Evans N. J., II, K\"ompe C., and Walmsley C. M. (1993) 
      {\it Astrophys.\ J., 404}, 232-246.

\refs Zucconi A., Walmsley C. M., and Galli D. (2001) {\it Astron.\ 
      Astrophys., 376}, 650-662.

\end{document}